\documentclass[12pt]{article}
\usepackage{mathrsfs}
\usepackage{authblk}
\usepackage{tipa}
\usepackage[capposition=below]{floatrow}
\usepackage{graphics}
\usepackage[svgnames]{xcolor}
\usepackage{amsmath}
\usepackage{amssymb}
\usepackage{bm}
\usepackage{amsfonts}
\usepackage{ntheorem}
\usepackage[colorlinks=true, linkcolor=blue!60!black, citecolor=green!50!black, linktoc=all]{hyperref}
\definecolor{darkgreen}{rgb}{0,0.5,0}
\definecolor{darkblue}{rgb}{0,0,0.6}
\definecolor{purple}{rgb}{0.4,0.15,0.21}
\numberwithin{equation}{section}
\usepackage{feynmf}
\usepackage{extarrows}
\usepackage{slashed}
\usepackage{graphicx}
\usepackage[page,title,titletoc,header]{appendix}
\usepackage{indentfirst}
\usepackage{color}
\author[]{Xiao Xiao\thanks{xx2146@columbia.edu}}
\affil[]{\normalsize{\emph{Physics Department and Institute for Strings, Cosmology, and Astroparticle Physics,}}\\
\emph{Columbia University, New York, NY 10027, USA}}
\begin{document}
\title{\large{\textbf{Holographic Representation of Local Operators In De Sitter Space}}}
\date{}
\maketitle
\abstract{Assuming the existence of the dS/CFT correspondence, we construct local scalar fields with $m^2>\left( \frac{d}{2} \right)^2$ in de Sitter space by smearing over conformal field theory operators on the future/past boundary. To maintain bulk microcausality and recover the bulk Wightman function in the Euclidean vacuum, the smearing prescription must involve two sets of single--trace operators with dimensions $\Delta$ and $d-\Delta$. Thus the local operator prescription in de Sitter space differs from the analytic continuation from the prescription in anti--de Sitter space. Pushing a local operator in the global patch to future/past infinity is shown to lead to an operator relation between single--trace operators in conformal field theories at $\mathcal{I}^\pm$, which can be interpreted as a basis transformation, also identified as the relation between an operator in CFT and its shadow operator. Construction of spin$-s$ gauge field operators is discussed, it is shown that the construction of higher spin gauge fields in de Sitter space is equivalent to constructing scalar fields with specific values of mass parameter $m^2<\left( \frac{d}{2} \right)^2$. An acausal higher spin bulk operator which matches onto boundary higher spin current is constructed. Implementation of the scalar operator constructions in AdS and dS with embedding formalism is briefly described.}

\clearpage

\tableofcontents

\numberwithin{equation}{section}
%%%%%%%%%%%%%%%%%%%%%%%
\section{Introduction}%
%%%%%%%%%%%%%%%%%%%%%%%
Gauge/gravity duality\cite{Malda}, which equates a theory of quantum gravity to a quantum field theory in one lower dimension, has provided a deeper understanding of both non--perturbative string theories and conformal field theories, and also finds applications in different areas such as nuclear physics and condense matter physics.

Despite the tremendous progress in the area of holographic duality, some basic questions regarding bulk locality remain to be clarified. Recently attention has been focused on sub--AdS locality \cite{HPPS}\cite{FK}---locality of physics within the AdS radius, which might help understanding the recent puzzles regarding black holes \cite{fire}. It is well--known that in order to be dual to weakly--coupled gravity in the form of a local field theory in AdS, a conformal field theory must have a large number of colors as well as being strongly coupled. The operator dictionary of AdS/CFT \cite{Wit}\cite{GKP}\cite{BDH} can be understood as a series of claims about locality in the near--boundary region of AdS. There are two kinds of operator dictionaries in AdS/CFT. One of them is the GKPW dictionary\cite{Wit}\cite{GKP} which identifies the boundary condition for a non--normalizable mode in AdS space as the coupling of a deformation to the boundary CFT, and the boundary correlation functions are obtained by differentiating this coupling to the partition function of bulk gravity. On the other hand, the BDHM dictionary \cite{BDH} identifies the boundary condition for a normalizable mode as an operator in the un--deformed CFT, and then CFT correlation functions are recovered by extrapolating the bulk quantum gravity correlation functions to the boundary. In both cases, there is a one--one correspondence between a local operator in the bulk and a local operator on the boundary. 

While the dictionary is well--defined in the limit that the bulk operator approaches the boundary, the story for an operator probing deeper inside the space is less transparent---such an operator corresponds to non--local operators on the boundary and the property of microcausality is not manifest. There are several approaches towards understanding this ``sub--AdS locality'' issue including the conformal bootstrap \cite{HPPS} and the use of Mellin representation of CFT correlation functions \cite{FK}. Constraints on operator dimensions and behavior of Mellin amplitudes are conjectured. In \cite{HPPS}, the authors count the constraints arising from the OPE, conformal invariance and the bootstrap conditions for large$-N$ conformal field theories in $d=2$ and $d=4$, and match the number of solutions to the constraints to the counting of quartic bulk local interactions. In \cite{FK}, CFT correlation functions are formulated as scattering amplitudes in AdS space, with the help of Mellin transform. It is demonstrated that to have local interactions in the AdS bulk, the Mellin amplitudes of the CFT should grow no faster than a power of the Mellin space coordinate $\delta$, in the limit that $\delta$ is large. In this paper, we focus on another approach which starts from microcausality and explicitly construct local operators from CFT data. The particular construction we are describing was developed in anti--de Sitter space by several authors \cite{HKLL}\cite{KLL}\cite{HMPS}, and recently further developed to describe the interior of eternal black hole in AdS space\cite{PR}, for exploring the ``firewall'' problem\cite{fire}. In this paper, we develop parallely the construction to local operators in de Sitter space, at the level of $\left( \frac{1}{N} \right)^0$ (two-point function), in the context of the de Sitter/CFT correspondence \cite{dSCFT}.

It is still not completely clear whether quantum gravity in de Sitter space can be described holographically: there are several proposals for such a correspondence, including dS/CFT\cite{dSCFT}, dS/dS\cite{dSdS} and static patch solipsism\cite{soli}. Among these dS/CFT seems to be the simplest extension of AdS/CFT to de Sitter space in the sense that quantities like CFT correlation functions and bulk wavefunction can be related to AdS case via analytic continuation, and the bulk de Sitter isometries match nicely to the conformal symmetry of the Euclidean theory at the future or past boundary. Recently there is a realization of dS/CFT proposed in the context of higher spin gravity\cite{AHS}; namely the Vasiliev theory\cite{Va} in de Sitter space is conjectured to be dual to critical or free Euclidean Sp(N) model with anti--commuting scalars. With nice analog to AdS/CFT as well as certain proposed realization though, the idea of dS/CFT suffers from several problems \cite{troub}\cite{Musing}: the CFT correlators are not observables for any observer in de Sitter space, rather they are ``meta--observables''\footnote{Actually we are meta--observers for the near de Sitter geometry during inflation \cite{NGau}, the CMB correlation functions are ``meta--observables'' for the observers in an inflating universe. CMB is observable to us because after inflation the universe exits the near--de Sitter phase and the CMB photons fall into causal contact with us.}, gravity is not decoupled, the dual field theory is non--unitary and it is hard to see how bulk unitarity arises. Also de Sitter future infinity may be spoiled by bubble nucleation and a boundary CFT may not exist at all.  

In this paper we will not get into any of these subtleties and just assume the existence of the dS/CFT correspondence, and try to construct local bulk observables from the boundary CFT data. We work in both the flat patch and the global patch of de Sitter space. In contrast to what happens in the case of AdS/CFT, at the level of two--point functions, a local operator in de Sitter space is shown to be constructed from two sets of single--trace operators in the boundary CFT, with dimension $\Delta$ and $d-\Delta$ respectively. The observation that an operator in de Sitter is dual to two operators in CFT is not itself new, having already been pointed out by the original dS/CFT paper \cite{dSCFT} by looking at the bulk correlation function in the limit that the bulk operators approach the boundary. In the paper by Harlow and Stanford\cite{HS}, the GKPW (differentiating) and BDHM (extrapolating) dictionaries in de Sitter space are shown to be inequivalent---while the ``differentiating'' dictionary gives correlators with a single scaling dimension, the ``extrapolating'' dictionary gives correlators with two different near--boundary behaviors. The construction of de Sitter local operators, according to the knowledge of the author, is new, and helps clarify the understanding of how the bulk observables of de Sitter space emerge from a lower dimensional space, as well as clarifying the difference between dS/CFT and the analytic continuation of AdS/CFT. We also generalize the construction to gauge fields with integer spin $s$. It is shown that the construction for a spin$-s$ gauge field in de Sitter space can be identified with a construction of scalar fields with $m^2<\left( \frac{d}{2} \right)^2$ and a bulk operator that matches with a spin$-s$ boundary current is explicitly constructed.

The structure of the paper is as follows: in section 2 we review the construction of local bulk operators in anti--de Sitter space. In section 3 we see that the analytic continuation of the AdS prescription to de Sitter space leads to acausal operators. In section 4 we describe the construction in the flat patch of de Sitter space, for scalars with mass parameter $m^2>\left(\frac{d}{2} \right)^2$ and recover the bulk Wightman function. Also we perform the construction in the global patch and derive a relation between operators on $\mathcal{I}^\pm$. We then discuss the extension of the construction for gauge fields with integer spin, and it is shown that for a local bulk gauge field, the construction is equivalent to the construction for scalar fields with mass parameter $m^2<\left( \frac{d}{2} \right)^2$. In the last part of the section we briefly describe the construction of AdS and dS scalar operators in the language of the embedding formalism. In section 5 we discuss the implications of the construction and possible open problems.

%%%%%%%%%%%%%%%%%%%%%%%%%%%%%%%%%%%%%%%%%%%%%%%%%%%%%%%%%%%%%%%%%%%%%
\section{Construction of a Local Scalar Field in Anti--de Sitter Space}%
%%%%%%%%%%%%%%%%%%%%%%%%%%%%%%%%%%%%%%%%%%%%%%%%%%%%%%%%%%%%%%%%%%%%%

In this section we look at how a bulk scalar field in an empty anti--de Sitter space emerges from a conformal field theory. We briefly review the construction in AdS space following\cite{HKLL}\cite{KLL}\cite{HMPS}. We work in the Poincar{\'e} patch, which has a direct analytic continuation to the flat slicing of de Sitter space.

There are two approaches leading to the same result. One is based on solving a space--like Cauchy problem and uses Green's function to express the local field, while the other starts from summing the normalizable modes in the bulk. In the Green's function approach, one first solves for the Green's function in AdS space
\begin{equation}
  \left( \Box'-m^2\right)G(z,x|z',x')=\frac{1}{\sqrt{-g}}\delta^{d}(x-x')\delta(z-z')
  \label{}
\end{equation}
Then from Green's theorem we have a bulk field expressed as
\begin{equation}
  \Phi(z,x)=\int_{z'\rightarrow 0}d^dx'\sqrt{-g'}\left( G(z,x;z',x')\partial_{z'}\Phi(z',x')-\Phi(z',x')\partial_{z'}G(z,x;z',x') \right)
  \label{}
\end{equation}
where for $\Phi(z,x)$ we just choose a single fall--off behavior near the boundary, as $z\rightarrow 0$
\begin{equation}
  \Phi(z,x)\sim z^\Delta \mathcal{O}(x)
  \label{}
\end{equation}
which corresponds to a normalizable solution to the bulk equation.

\begin{figure}[tbp]
   \centering
   \includegraphics[width=6cm]{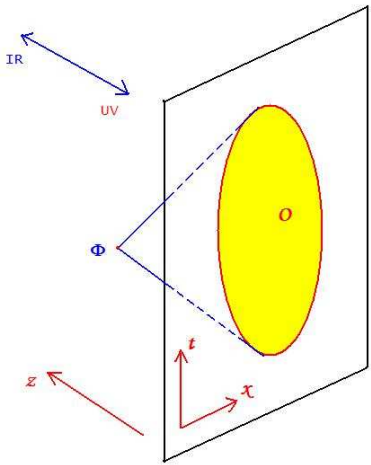}
   \caption{Construction of A Local Observable in Anti--de Sitter space}
   \label{AdS3}
 \end{figure}

Pushing $z'$ to the boundary and using the Green's function, finally we get
\begin{equation}
  \Phi(z,x)=\int d^dx' K\left( z,x|x' \right)\mathcal{O}(x')
  \label{}
\end{equation}
with $K(z,x|x')$ being a function which behaves like $z^{d-\Delta}$ when $z$ approaches zero. We call it the ``smearing function'', for it smears over the operators in a certain region in CFT, defining a non--local operator in CFT as a local operator in the bulk. In the Poincar{\'e} patch it is
\begin{equation}
  \boxed{  K(z,x|x')=c_{d,\Delta}\left( \frac{z^2+\left( x-x' \right)^2}{z} \right)^{\Delta-d}\Theta\left( z-|x-x'| \right)}
  \label{}
\end{equation}
The domain of integration on the boundary is finite and within the intersection with the bulk lightcone from the bulk operator, as shown in figure (\ref{AdS3}). Though it looks like an unconventional Cauchy problem---the ``initial data'' are spacelike separated from the bulk point, the result is causal: the commutator between two bulk operators constructed in this way vanishes when they are spacelike separated, to order $N^0$ in the large--N expansion. When considering interactions, the commutator turns on in three--point functions, which is at the order $N^{-1}$, but this can be cured by including multi--trace operators in the smearing prescription \cite{KLL}. Schematically we have:
\begin{equation}
  \Phi(z,x)=\int d^dx'K(z,x|x')\mathcal{O}(x')+\sum_{i}\int d^dx'K_{\Delta_i}\left( z,x|x' \right)\mathcal{O}_{\Delta_i}(x')
  \label{}
\end{equation}
with multi--trace operators $\mathcal{O}_{\Delta_i}$ such as $\mathcal{O}^2$. The commutator at the order of $N^{-1}$ is then cancelled by the contributions from the multi--trace operators. With concrete demonstration at the order of $N^{-1}$ \cite{KLL}, the procedure is conjectured to work order--by--order in the large--N expansion, and by adding multi--trace operators one can construct local operators in AdS to any order of $\frac{1}{N}$. The construction is believed to break down away from the large--N limit, where bulk gravity fluctuates and the notion of microcausality as well as the notion of the background spacetime itself break down; one would not expect to define local observables in a full--fledged gravity theory \cite{dWit}

The second approach---summing over modes---is more transparent for seeing why the result is causal. One starts by solving the free field equation in AdS and ends up with two independent solutions, which in the example of $AdS_2$ are
\begin{equation}
  \Phi_{\omega}(z)\sim C_1\sqrt{z}J_{\Delta-\frac{1}{2}}(z\omega)+C_2\sqrt{z}Y_{\Delta-\frac{1}{2}}(\omega z)
  \label{}
\end{equation}
where $J_\nu$ and $Y_\nu$ are Bessel functions. Only the part proportional to $J_\nu$ is normalizable, so we just keep this branch of the solutions and sum over it. 

Summing over the normalizable modes, including both positive frequencies and negative frequencies, we get
\begin{equation}
  \Phi(z,t)=\int^\infty_0 d\omega\left( a_\omega e^{-i\omega t}+a_\omega^\dag e^{i\omega t} \right)\sqrt{z}J_{\Delta-\frac{1}{2}}(\omega z)
  \label{}
\end{equation}
which will recover the same result obtained by the Green's function approach. 

Here we can easily see why the result should be causal: although we just choose one of the fall--off behaviors in the $z$--direction, we are keeping both positive frequencies and negative frequencies in the time direction. This is crucial for ensuring microcausality, just as in flat spacetime. Thus we can express a local operator in $AdS_{d+1}$ space in terms of the $CFT_d$ operators inside its spatial lightcone, and these bulk operators satisfy microcausality. 

We see that a local operator inside AdS space emerges as a non--local operator on the boundary. From the boundary point of view, the AdS coordinates $t$ and $z$ are just parameters defining the non--local operator in the CFT. The duality between bulk and boundary physics ensures that this particular non--local operator in CFT satisfies a free field equation in a higher dimension as well as being local in the sense of a higher dimensional microcausality. We should emphasize that the map between boundary and bulk here depends on the state of the boundary CFT \cite{PR}, which maps to a certain bulk background geometry. The smearing function---and thus the construction---is made refering to a certain background. In this case it is empty AdS space, which is dual to the vacuum of a zero--temperature CFT. In a more general background, for instance, with a black hole sitting in the bulk, the construction would be different. 

%%%%%%%%%%%%%%%%%%%%%%%%%%%%%%%%%%%%%%%%%%%%%%%%%%%%%%%%%%
\section{Analytic Continuation and Operator Dictionaries}%
%%%%%%%%%%%%%%%%%%%%%%%%%%%%%%%%%%%%%%%%%%%%%%%%%%%%%%%%%%

We will see that though de Sitter space and the anti--de Sitter space are related to each other via analytic continuation, the analytic continuation of the AdS smearing prescription above to de Sitter space does not give causal correlation functions.  As a starting point one can analytically continue the AdS Poincar{\'e} patch to de Sitter flat slicing via
\begin{equation}
  z\rightarrow \eta~,~t\rightarrow t~,~x^i\rightarrow ix^i~ ,~R_{AdS}=iR_{dS}
  \label{}
\end{equation}
and get
\begin{equation}
  ds^2=\frac{-d\eta^2+d\textbf{x}^2}{\eta^2}
  \label{}
\end{equation}
with $t$ treated as one of the spatial coordinates in de Sitter space \footnote{Here we just set $R_{dS}$ to one.}
.

Thus if one does the analytic continuation to the prescription introduced in the section above, a field operator in de Sitter space is then expressed as an integral defined on the past or future boundary. The domain of integration for smearing is in the past/future light cone of the bulk point, and it is now a standard Cauchy problem to express the bulk point in terms of boundary operators as evolving the initial conditions using the retarded Green's function in de Sitter space. 

However this cannot be what we aim for. First, it violates microcausality: after the analytic continuation, the spatial lightcone in AdS becomes time--like, and the bulk operator will now commute with the operators inside its own time--like lightcone and fails to commute with the ones outside, which is not the right behavior for being causal. One can see the reason why this happens---in AdS smearing we just sum one set of the modes which are the normalizable ones into the local operator. Continuing to de Sitter space the $z$ direction becomes the time direction and keeping only one set of modes in this direction turns into keeping either positive or negative frequency modes, which spoils microcausality. In AdS we do not have this problem because we go from the bulk to the boundary in a spatial direction, and we can still keep both positive frequency and negative frequency modes in the time direction while sticking to just normalizable modes in $z$ direction.

Second, from the smearing prescription above one recovers the correct AdS bulk correlation functions, but as has been pointed out by several authors \cite{HS}\cite{BMS}\cite{dSCFT}, the analytic continuation of AdS correlators would not give the correlation function in any de Sitter invariant vacuum. In paper \cite{HS} the authors used the language of the holographic renormalization group \cite{HP} to clarify this point. They claim that in de Sitter space the ``GKPW dictionary'' and the ``BDHM dictionary'' are not equivalent, though the vacuum wavefunctions in AdS and dS are related by analytic continuation. The reason why this happens is the definitions of correlation functions in dS and AdS are not related to each other via analytic continuation. In the language of the holographic RG, one can define the correlation functions in AdS in the following way \cite{HS}: split the bulk path integral with a plane at $z=l$ and the path integrals in the UV side and IR side give UV and IR wavefunctions separately, and then one can insert operators on the plane and thus obtain a bulk correlation function
\begin{equation}
  \langle\tilde{\phi}(x_1,l)\dots\tilde{\phi}(x_n,l)\rangle_{AdS}=\int_{z=l} \mathcal{D}\tilde{\phi}\Psi_{IR}[\tilde{\phi}]\tilde{\phi}(x_1,l)\dots\tilde{\phi}(x_n,l)\Psi_{UV}[\tilde{\phi},\phi_0]
  \label{}
\end{equation}
where $\phi_0$ is the boundary condition for the path integral in $\Psi_{UV}$. One recovers the boundary correlation function by taking the limit $l\rightarrow 0$, and it agrees with the result one gets by differentiating with the boundary condition $\phi_0$ \cite{HS}. The $\Psi_{IR}$ is shown to be related to the Hartle--Hawking vacuum in de Sitter space $\Psi_{HH}$ via analytic continuation \cite{HS}; however if we analytically continue the definition of the correlation function to de Sitter space one gets something peculiar: taking the future wedge of flat slicing, the $\Psi_{UV}$ is now a wavefunction in the later stage of the universe and we call it $\Psi_L$, and $\Psi_{IR}$ is defined in an earlier period and we call it $\Psi_{E}$. Then the analytically continued correlation function is defined as
\begin{equation}
  \langle\tilde{\phi}(x_1,\eta)\dots\tilde{\phi}(x_n,\eta)\rangle_{dS}=\int_{\eta} \mathcal{D}\tilde{\phi}\Psi_{E}[\tilde{\phi}]\tilde{\phi}(x_1,\eta)\dots\tilde{\phi}(x_n,\eta)\Psi_{L}[\tilde{\phi},\phi_0]
  \label{}
\end{equation}
This is different from how one computes correlation functions in de Sitter space, or in a more generic FRW cosmology. With this definition, in order to compute the correlation functions at a certain time $\eta$ it is not enough to know the earlier stage evolution of the wavefunction, but as well the later stage, this is not what one would do in cosmology since to compute the correlation functions of temperature fluctuations in cosmic microwave background we don't have to know the wavefunction of universe during the subsequent structure formation. Also, fixing a certain boundary condition at the future infinity is manifestly acausal \cite{ANS}. The radiation fails to pass through the future infinity, it will be reflected back into the past. This acausal behavior will manifest itself as the breakdown of microcausality: operators on a single spatial slice fail to commute. The way to define the correlation function in de Sitter and in more generic cosmology should just involve the Hartle--Hawking wavefunction and its complex conjugate, and corresponds to an in--in path integral:

\begin{equation}
\langle\Psi|\tilde{\phi}(x_1,\eta)\dots\tilde{\phi}(x_n,\eta)|\Psi\rangle_{dS,FRW}=\int_{\eta} \mathcal{D}\tilde{\phi}\Psi_{E}^*[\tilde{\phi}]\tilde{\phi}(x_1,\eta)\dots\tilde{\phi}(x_n,\eta)\Psi_{E}[\tilde{\phi}]
  \label{}
\end{equation}
where $\eta$ is a certain spatial slice on which we compute correlation functions and compare with data, such as the last scattering surface of CMB photons in our universe, and $\Psi_E$ refers to both ``a wavefunction at early time'' and ``a wavefunction of the universe in the Euclidean (Hartle--Hawking) vacuum''. Here one no longer specifies the boundary condition at the future boundary. This is a natural definition of expectation values under the Born rule, and it is clearly different from the analytic continuation from AdS. Also this definition obeys microcausality---the spacelike separated operators commute inside the correlation functions and timelike separated ones do not commute. The simplest one of this type of correlation functions is the Wightman function for a free scalar field in de Sitter space. Thus a construction of a de Sitter bulk operator that computes de Sitter cosmology should reproduce the Wightman function, and it should also contain both positive and negative frequency modes in de Sitter space in order to ensure causality.

%%%%%%%%%%%%%%%%%%%%%%%%%%%%%%%%%
\section{De Sitter Construction}%
%%%%%%%%%%%%%%%%%%%%%%%%%%%%%%%%%

\subsection{Flat Slicing}

Now we look at how a local scalar operator with mass $m^2>\left( \frac{d}{2} \right)^2$ in de Sitter space is constructed with a CFT located at the boundary. In the AdS construction the boundary is timelike, and the extra direction emerges from the boundary as a spatial direction. In de Sitter space, the boundaries are located at future and past infinity, which are spacelike boundaries, so what emerges from the CFT is the bulk time. From the boundary point of view the bulk time $\eta$ appears as a parameter in the definition of non--local CFT operators. As we will see, a local bulk operator that is far from the boundary will be highly non--local from the CFT point of view. 

In this subsection we work in the flat patch of de Sitter space, which covers only a half of the global geometry. One can either choose the past wedge to work on, or the future wedge, and the boundary CFT will live on $\mathcal{I}^-$ or $\mathcal{I}^+$ respectively. Here for the moment we choose the past wedge. The construction in the global patch of de Sitter space is left to the next subsection.

We have seen that a construction prescription for local operators in de Sitter space should involve modes with both positive and negative frequencies, corresponding to ``normalizable'' and ``non-normalizable'' behaviors in AdS. Here we define
\begin{equation}
  \Delta=\frac{d}{2}+i\sqrt{m^2-\left( \frac{d}{2} \right)^2}
  \label{}
\end{equation}
and near the boundary a positive/negative frequency mode has behavior
\begin{equation}
  \Phi_+\left( \eta\rightarrow 0 \right)\sim \eta^\Delta\mathcal{O}_+~,~\Phi_-\left( \eta\rightarrow 0 \right)\sim\eta^{d-\Delta}\mathcal{O}_-.
  \label{}
\end{equation}
where $\mathcal{O}_\pm$ are single--trace operators in the boundary CFT, with scaling dimensions $\Delta$ and $d-\Delta$ respectively.

For the case of interest here, since $m^2-\left( \frac{d}{2} \right)^2$ is positive, near the boundary both $\Phi_+$ and $\Phi_-$ are damped by the same factor $\eta^{\frac{d}{2}}$ and oscillate with frequency $\sqrt{m^2-\left( \frac{d}{2} \right)^2}$. If $m^2<\left( \frac{d}{2} \right)^2$ then the two modes fall at different rates near the boundary and do not oscillate.

According to the reasoning in the section above, a causal operator should have both components, schematically:
\begin{equation}
  \Phi(\eta\rightarrow 0)\sim A\eta^\Delta\mathcal{O}_++B\eta^{d-\Delta}\mathcal{O}_-
  \label{}
\end{equation}
With a certain linear combination, one can reproduce the Wightman function in the Euclidean vacuum. 

To construct the bulk operator, we evolve the initial data at $\mathcal{I}^-$ with the retarded Green's function, which is
\begin{equation}
  G_{ret}|_{\eta'\rightarrow 0}\approx c_{\Delta,d}(-\sigma-i\epsilon)^{\Delta-d}+c_{\Delta,d}^*(-\sigma-i\epsilon)^{-\Delta}-c.c.
  \label{}
\end{equation}
in the limit that $\eta'\rightarrow 0$. Here 
\begin{equation}
  \sigma=\frac{\eta^2+\eta'^2-\left( \textbf{x}-\textbf{x}' \right)^2}{2\eta\eta'}
  \label{}
\end{equation}
is a de Sitter invariant distance and 
\begin{equation}
  c_{\Delta,d}=\frac{\Gamma(2\Delta-d)\Gamma(d-\Delta)}{2^{\Delta-d}(4\pi)^{\frac{d+1}{2}}\Gamma(\Delta-\frac{d-1}{2})}
  \label{}
\end{equation}
The bulk operator is constructed by evolving an operator near the boundary:
\begin{equation}
  \Phi(\eta,\textbf{x})=\int_{|\textbf{x}'|<\eta}d^dx'\left( \frac{1}{\eta'} \right)^{d-1}\left( G_{ret}(\eta,\textbf{x};\eta',\textbf{x}')\partial_{\eta'}\Phi(\eta',\textbf{x}')-\Phi(\eta',\textbf{x}')\partial_{\eta'}G_{ret}(\eta,\textbf{x};\eta',\textbf{x}') \right)
  \label{}
\end{equation}
where 
\begin{equation}
  \Phi(\eta',\textbf{x}')\sim A\left(\eta'\right)^\Delta\mathcal{O}_+(\textbf{x}')+B\left(\eta'\right)^{d-\Delta}\mathcal{O}_-\left( \textbf{x}' \right)
  \label{}
\end{equation}
 By keeping both sets of operators, we are keeping both the positive and negative frequency parts of the solution \footnote{In evaluating the equation above, for the positive or negative frequency component the retarded Green's function we have both terms that are proportional to $\sigma^{\Delta-d}\mathcal{O}_\pm$ and $\sigma^{-\Delta}\mathcal{O}_\pm$, but in the limit $\eta'\rightarrow 0$ only $\sigma^{\Delta-d}\mathcal{O}_+$ and $\sigma^{-\Delta}\mathcal{O}_-$ survive because the other contributions oscillate quickly and go to zero as $\eta'$ approaches zero. The details are presented in Appendix B.}.

Evaluating the integrand, we have the local operator in de Sitter space expressed as
\begin{equation}
  \Phi(\eta,\textbf{x})=A_{\Delta,d}\int_{|\textbf{x}'|<\eta}d^dx'\left( \frac{\eta^2-\textbf{x}'^2}{\eta} \right)^{\Delta-d}\mathcal{O}_+(\textbf{x}+\textbf{x}')+B_{\Delta,d}\int_{|\textbf{x}'|<\eta}d^dx'\left( \frac{\eta^2-\textbf{x}^2}{\eta} \right)^{-\Delta}\mathcal{O}_-(\textbf{x}+\textbf{x}')
  \label{smear}
\end{equation}
where $A_{\Delta,d}$ and $B_{\Delta,d}$ are certain coefficients. Here for a moment we keep them free, since in principle we can rescale the boundary operators and change the coefficients of the two--point functions, as a marginal deformation to the boundary CFT. This freedom of rescaling the operators, as well as the freedom of choosing a certain linear combination of two modes with different fall--off behaviors, enables us to keep $A_{\Delta,d}$ and $B_{\Delta,d}$ free for a moment. Here we will finally fix them by demanding that the correlation function of $\Phi$ recover the Wightman function in the Euclidean vacuum. This means that the choice of the coefficients is state--dependent: for other de Sitter invariant vacua such as ``$\alpha$--vacua'', we should have different prescriptions in order to recover Wightman functions.

 \begin{figure}[tbp]
   \centering
   \includegraphics[width=12cm]{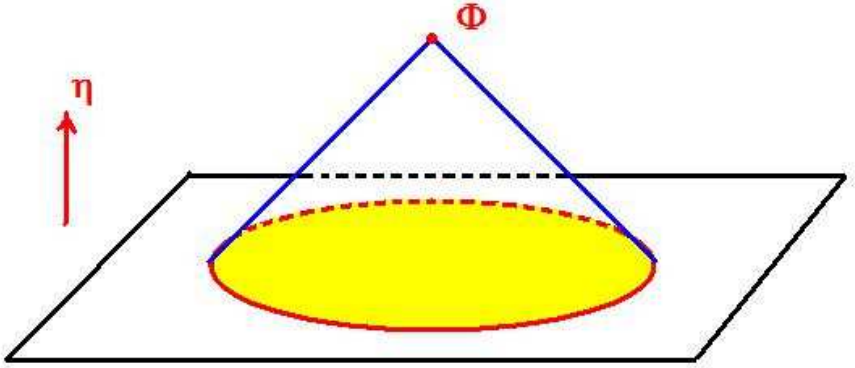}
   \caption{Construction in the Flat Patch of De Sitter Space}
   \label{ds3}
 \end{figure}
Having a prescription, we would like to calculate the bulk two--point function and compare with the two--point bulk Wightman function in the limit that one of the bulk points approaches the boundary. The Euclidean vacuum Wightman function in de Sitter space is \cite{BMS}
\begin{equation}
  G_E\left( x,x' \right)=\frac{\Gamma(\Delta)\Gamma(d-\Delta)}{(4\pi)^{\frac{d+1}{2}}\Gamma(\frac{d+1}{2})}F\left( \Delta,d-\Delta,\frac{d+1}{2},\frac{1+\sigma}{2} \right)
  \label{wight}
\end{equation}
where $F(\alpha,\beta,\gamma,x)$ is the hypergeometric function $~_2F_1$, and $\sigma$ is the de Sitter invariant distance defined beforehand.

When one of the bulk points $x'$ approaches the boundary $\eta'\rightarrow 0$, the fourth argument of the hypergeometric function grows large and is dominated by $\sigma$
\begin{equation}
  \frac{1+\sigma}{2}\sim\frac{\sigma}{2}\sim\frac{\eta^2-(\textbf{x}-\textbf{x}')^2}{4\eta\eta'}
  \label{}
\end{equation}
For convenience we can set $\textbf{x}'$ to zero. In this limit we have

 \begin{equation}
   G_E(\eta,\textbf{x};\eta'\sim0,\textbf{x}'=0)\rightarrow \frac{\Gamma(\Delta)\Gamma(d-2\Delta)}{(4\pi)^{\frac{d+1}{2}}\Gamma(\frac{d+1}{2}-\Delta)}\left(- \frac{4\eta\eta'}{\eta^2-\textbf{x}^2} \right)^{\Delta}+\frac{\Gamma(2\Delta-d)\Gamma(d-\Delta)}{(4\pi)^{\frac{d+1}{2}}\Gamma(\Delta-\frac{d-1}{2})}\left( -\frac{4\eta\eta'}{\eta^2-\textbf{x}^2} \right)^{d-\Delta}
   \label{}
 \end{equation}
 As expected, the Wightman function has two components with dimensions $\Delta$ and $d-\Delta$. Next we want to reproduce it from the smearing formula (\ref{smear}).

 Here we would like to normalize the boundary two--point functions so that we have
 \begin{equation}
   G_E\left( \eta\rightarrow 0,\textbf{x};\eta'\rightarrow 0,\textbf{x}'=0 \right)\rightarrow (\eta\eta')^\Delta D_+\left(\textbf{x}  \right)+\left( \eta\eta' \right)^{d-\Delta}D_-\left( \textbf{x} \right)
   \label{}
 \end{equation}
 where $D_\pm$ are the boundary CFT correlation functions which we take to be
 \begin{align}
   & D_+(\textbf{x})= \frac{2^{2\Delta}\Gamma(\Delta)\Gamma(d-2\Delta)}{(4\pi)^{\frac{d+1}{2}}\Gamma(\frac{d+1}{2}-\Delta)}\left( \frac{1}{\textbf{x}^2} \right)^\Delta\\
   & D_-(\textbf{x})= \frac{2^{2(d-\Delta)}\Gamma(2\Delta-d)\Gamma(d-\Delta)}{(4\pi)^{\frac{d+1}{2}}\Gamma(\Delta-\frac{d-1}{2})}\left( \frac{1}{\textbf{x}^2} \right)^{d-\Delta}
   \label{}
 \end{align}
 Taking the smearing formula (\ref{smear}) and computing the correlation function between the bulk operator and an operator near the boundary, we have

 \begin{equation}
   \begin{split}
   & \langle\Phi(\eta,\textbf{x})\Phi(\eta'\rightarrow 0,0)\rangle=
   A_{\Delta,d}\int_{|\textbf{x}'|<\eta}d^dx'\left( \frac{\eta^2-\textbf{x}'^2}{\eta} \right)^{\Delta-d}\eta'^\Delta\langle\mathcal{O}_+(\textbf{x}+\textbf{x}')\mathcal{O}_+(0)\rangle\\
   &   +B_{\Delta,d}\int_{|\textbf{x}'|<\eta}d^dx'\left( \frac{\eta^2-\textbf{x}'^2}{\eta} \right)^{-\Delta}\eta'^{d-\Delta}\langle\mathcal{O}_-(\textbf{x}+\textbf{x}')\mathcal{O}_-(0)\rangle
 \end{split}
 \end{equation}
 With the boundary correlator of the operators $\mathcal{O}_\pm$:
 \begin{equation}
   \langle\mathcal{O}_+(\textbf{x})\mathcal{O}_+(0)\rangle=D_+(\textbf{x})~,~\langle\mathcal{O}_-(\textbf{x})\mathcal{O}_-(0)\rangle=D_-(\textbf{x})~,~\langle\mathcal{O}_+(\textbf{x})\mathcal{O}_-(0)\rangle=0
   \label{}
 \end{equation}
 we obtain

 \begin{equation}
   \begin{split}
   & \langle\Phi(\eta,\textbf{x})\Phi(\eta'\rightarrow 0,0)\rangle=
   A_{\Delta,d}\frac{2^{2\Delta}\Gamma(\Delta)\Gamma(d-2\Delta)}{(4\pi)^{\frac{d+1}{2}}\Gamma(\frac{d+1}{2}-\Delta)}\int_{|\textbf{x}'|<\eta}d^dx'\left( \frac{\eta^2-\textbf{x}'^2}{\eta} \right)^{\Delta-d}\eta'^\Delta\frac{1}{\left( \textbf{x}+\textbf{x}' \right)^{2\Delta}}\\
   &   +B_{\Delta,d}\frac{2^{2(d-\Delta)}\Gamma(2\Delta-d)\Gamma(d-\Delta)}{(4\pi)^{\frac{d+1}{2}}\Gamma(\Delta-\frac{d-1}{2})}\int_{|\textbf{x}'|<\eta}d^dx'\left( \frac{\eta^2-\textbf{x}'^2}{\eta} \right)^{-\Delta}\eta'^{d-\Delta}\frac{1}{\left( \textbf{x}+\textbf{x}' \right)^{2(d-\Delta)}}
 \end{split}
 \end{equation}
 After evaluating the integrals we end up with the result

 \begin{equation}
   \begin{split}
   &  \langle\Phi(\eta,\textbf{x})\Phi(\eta'\rightarrow 0,0)\rangle=\\
   & A_{\Delta,d}\frac{\pi^{\frac{d}{2}}\Gamma(\Delta-d+1)}{\Gamma(\Delta-\frac{d}{2}+1)} \frac{2^{2\Delta}\Gamma(\Delta)\Gamma(d-2\Delta)}{(4\pi)^{\frac{d+1}{2}}\Gamma(\frac{d+1}{2}-\Delta)}(-1)^\Delta\left( \frac{\eta\eta'}{\eta^2-\textbf{x}^2} \right)^\Delta+\\
&   B_{\Delta,d} \frac{\pi^{\frac{d}{2}}\Gamma(1-\Delta)}{\Gamma(\frac{d}{2}-\Delta+1)}\frac{2^{2(d-\Delta)}\Gamma(2\Delta-d)\Gamma(d-\Delta)}{(4\pi)^{\frac{d+1}{2}}\Gamma(\Delta-\frac{d-1}{2})}(-1)^{d-\Delta}\left( \frac{\eta\eta'}{\eta^2-\textbf{x}^2} \right)^{d-\Delta}
\end{split}
 \end{equation}
 We see that the coefficients

 \begin{align*}
   & A_{\Delta,d}=\frac{\Gamma(\Delta-\frac{d}{2}+1)}{\pi^{\frac{d}{2}}\Gamma(\Delta-d+1)}\\
   & B_{\Delta,d}=A_{d-\Delta,d}=\frac{\Gamma(\frac{d}{2}-\Delta+1)}{\pi^{\frac{d}{2}}\Gamma(1-\Delta)}
 \end{align*}
 give the two--point Wightman function in the Euclidean vacuum, in the limit that one of the bulk points approaches the boundary. It is not hard to show that the prescription also gives Wightman function away from this limit. Notice that each of the two terms in the smearing prescription can actually be obtained by analytically continuing from AdS space. Thus the Wightman function breaks into two pieces. One is for a scalar with $\Delta_+=\Delta=\frac{d}{2}+i\sqrt{m_{dS}^2-\left( \frac{d}{2} \right)^2}$ and the other for a scalar with $\Delta_-=d-\Delta_+$. The two--point functions for the $\Delta_+$ and $\Delta_-$ components are in the form:
 \begin{equation}
   G_\Delta(x,x')=\left( \frac{2}{1+\sigma} \right)^{\Delta}F\left( \Delta,\Delta-\frac{d-1}{2},2\Delta-d+1,\frac{2}{1+\sigma} \right)
   \label{}
 \end{equation}
 with $\Delta=\Delta_\pm$.

 Summing the two pieces using the property of hypergeometric function introduced in Appendix. C, with the coefficients obtained in this section, we recover the Wightman function in de Sitter space with both points deep in the bulk.

 Now we can write the expression for a local operator in de Sitter space explicitly, with coefficients set by the Euclidean vacuum state:
 \begin{equation}\boxed{\begin{split}
  \Phi(\eta,\textbf{x})&=\frac{\Gamma(\Delta-\frac{d}{2}+1)}{\pi^{\frac{d}{2}}\Gamma(\Delta-d+1)}\int_{|\textbf{x}'|<\eta}d^dx'\left( \frac{\eta^2-\textbf{x}'^2}{\eta} \right)^{\Delta-d}\mathcal{O}_+(\textbf{x}+\textbf{x}')+\\
  &\frac{\Gamma(\frac{d}{2}-\Delta+1)}{\pi^{\frac{d}{2}}\Gamma(1-\Delta)}\int_{|\textbf{x}'|<\eta}d^dx'\left( \frac{\eta^2-\textbf{x}^2}{\eta} \right)^{-\Delta}\mathcal{O}_-(\textbf{x}+\textbf{x}')
\end{split}}
  \label{}
\end{equation}
This is our main result: to construct local operators in de Sitter space that probe and create particles in the Euclidean vacuum state, we start from the Wightman function in the bulk Euclidean vacuum $G_E(x,x')$, and construct the retarded propagator by taking the expectation value of the commutator $G_{ret}\equiv G_E(x,x')-G_E(x',x)$ which has support only inside the bulk time--like lightcone. This retarded propagator gives the smearing functions for CFT operators with coefficients $A_{\Delta,d}$ and $B_{\Delta,d}$, and using the constructed smearing function we can recover the Wightman function we started with. What we get is a representation of local bulk operator in terms of boundary CFT operators, in a certain vacuum state, the bulk operator is constructed with CFT data inside the past lightcone of the bulk point, as shown in figure (\ref{ds3}).

 Here one can also see that as opposed to the AdS case, to check microcausality one is no longer supposed to just compute the correlation function between a bulk operator and a single boundary operator. Considering for example $\mathcal{O}_+$, the result one gets in this way is the same as the one continued from AdS, and thus acausal. The reason why we shouldn't do this is clear: unlike the case for AdS, $\mathcal{O}_+$ or $\mathcal{O}_-$ alone no longer match smoothly onto any local bulk operator that approaches the boundary.

 The construction above in terms of CFT operators at past boundary $\mathcal{I}^-$ is not directly relevant to cosmology. In cosmology it is the flat FRW slicing of de Sitter space defined on the future wedge which is relevant. It describes the expansion phase of the universe. In the future wedge the bulk operators are constructed with CFT operators on $\mathcal{I}^+$, which seems unappealing because the ``retarded propagator'' is now propagating the boundary operators back in time. However in terms of the physical observables everything is causal: by rerunning the calculation we obtain the Wightman function in the future wedge, and the operators satisfy microcausality. From the point of view of the evolution of wavefunctions, it is more appealing to phrase the construction in the future wedge: to compute correlation functions at a late time $\eta\rightarrow 0$ one starts with the vacuum state defined on some spatial slice at earlier time $\eta\rightarrow -\infty$ and evolve forward in time, then compute the expectation value. In the above construction we chose $\mathcal{I}^-$ simply because it is more appealing from the perspective of retarded propagator. One can rerun everything we formulated above in the future wedge and get a local operator in the future wedge in terms of operators at $\mathcal{I}^+$.

 One may also question about the operator content: in a CFT, it seems we don't necessarily have operator content that is enough for the construction. Take the example of a scalar. It is totally possible that the theory only contains a dynamical scalar current $\mathcal{O}_+$ with dimension $\Delta$ but not one with dimension $d-\Delta$. However we can introduce such an operator by interpreting the coupling $\beta$ to $\mathcal{O}_+$ as the $\mathcal{O}_-$ that we need. This means that when doing the path integral for the CFT, we are not only integrating over the constituents that forming $\mathcal{O}_+$, but also the source \footnote{I thank Frederik Denef for pointing this out.}. This is legitimate as one can always treat the coupling as a multiplier and integrate over it. Furthermore, this is consistent with the fact that when computing expectation values in de Sitter space, we have to integrate over sources also. An example is to compute correlation functions at future infinity with the wavefunction of the universe $\Psi[g_{ij}]$. To obtain this wavefunction we do a path integral in de Sitter space using the Hartle--Hawking prescription, which is equivalent to coupling $g_{ij}$ as a source to the boundary stress tensor and integrating over the CFT field contents. When we want to compute correlation functions on the boundary, we also have to integrate over $g_{ij}$, which is a quantum number that corresponds to a certain classical configuration now, according to Born's rule. A CFT enlarged to include operators corresponding to sources then becomes a ``doubled CFT'' as discussed in \cite{HS}\cite{HSu}. In the next subsection, we can see that the second set of operators $\mathcal{O}_-$ also has a natural interpretation as the ``shadow operators'' in the CFT.

 Further, one may wonder if by writing down $\Phi\rightarrow A\eta^{\Delta}\mathcal{O}_++B\eta^{d-\Delta}\mathcal{O}_-$ near the boundary, we are imposing boundary conditions similar to what we do in anti--de Sitter space. From the perspective of path integration, we have seen that what produces the Wightman function in de Sitter space is an in--in type path integral which does not fix any boundary condition at the future or past boundary. Rather it specifies a particular vacuum state in which we calculate the expectation values. Indeed we are not fixing the boundary condition even we write down a schematic form of the operator near the boundary. The reason is that we are not fixing the coefficients: $\Phi$ can be any linear combination of the two components. What fixes a particular linear combination is the vacuum we choose: here we chose a particular A and B to recover the Wightman function in the Euclidean vacuum. Therefore there is no contradiction with the fact that the correlation functions are computed by an in--in path integral. 

 One further comment about the case of $m^2\leq\left( \frac{d}{2} \right)^2$ here. Our construction is done for the case $m^2>\left( \frac{d}{2} \right)^2$ and we have seen the positive and negative frequency modes join nicely into the Wightman function in the Euclidean vacuum. For $m^2<\left( \frac{d}{2} \right)^2$, apart from some specific values, one can continue our result trivially and obtain Wightman function for the light scalar. This corresponds to summing over bulk modes with near boundary behavior $\eta^\Delta$ and $\eta^{d-\Delta}$ with $\Delta=\frac{d}{2}+\sqrt{\left( \frac{d}{2} \right)^2-m^2}$. However in this case the modes have no oscillatory behavior---they just fall at different rates, thus the interpretation of positive--negative frequencies is not a good one despite the fact that we can construct the local bulk operator in the same way. For some specific values of $m^2$, the construction fails, as we will see in section $\ref{subsec:gauge}$, the reason is when $2\Delta-d$ is an integer---which happens when we have a light scalar with $m^2=-(s-2)(s+d-2)\leq\left( \frac{d}{2} \right)^2$, with $s$ here a positive integer which turns out to be the spin of a spin--$s$ gauge field---the Wightman function no longer split into two parts corresponding to complementary dimensions and there is a logarithmic term. Our construction formula has explicit singularities on such mass parameters.
 
 \subsection{Global Slicing}%
 
 The construction in the global patch is similar to the flat patch, but with new elements from having two boundaries. Now we can define conformal field theory operators separately at $\mathcal{I}^+$ and $\mathcal{I}^-$; for any local operator in the bulk, the CFT operators at $\mathcal{I}^+$ and $\mathcal{I}^-$ can be regarded as two different bases, which should be related to each other via a Bogoliubov transformation \cite{BMS}. In order to see the relation, one may construct a local field with CFT operators in $\mathcal{I}^-$ and push it to $\mathcal{I}^+$, or vice versa, thus getting the expression of an operator on $\mathcal{I}^+$ in terms of operators on $\mathcal{I}^-$. As a starting point, we first formulate the global patch smearing function.

 In the global patch we work in conformal time, with the metric
 \begin{equation}
   ds^2=\frac{1}{\cos^2\tau}\left( -d\tau^2+d\Omega_d^2 \right).
   \label{}
 \end{equation}
 Here the topology of the spacetime is $R\times S^d$ with the conformal time $\tau$ running from $-\frac{\pi}{2}$ to $\frac{\pi}{2}$.

 In these coordinates, the de Sitter invariant distance is expressed as
 \begin{equation}
   \sigma(x,x')=\frac{\cos(\Omega-\Omega')-\sin\tau\sin\tau'}{\cos\tau\cos\tau'}.
   \label{}
 \end{equation}

 As $x'$ goes to future boundary $\mathcal{I}^+$, $\tau'$ goes to $\frac{\pi}{2}$, and the regularized distance from a bulk point to the boundary point is
 \begin{equation}
   \sigma(x,x')\cos\tau'\sim \frac{\cos(\Omega-\Omega')-\sin\tau}{\cos\tau}
   \label{}
 \end{equation}
 Therefore the smearing functions that evolve future boundary operators back into the bulk will be proportional to
 \begin{align*}
   & K_+^{\mathcal{I}^+}(\tau,\Omega|\Omega')\sim\left( \frac{\cos(\Omega-\Omega')-\sin\tau}{\cos\tau} \right)^{\Delta-d}\\
   & K_-^{\mathcal{I}^+}(\tau,\Omega|\Omega')\sim\left( \frac{\cos(\Omega-\Omega')-\sin\tau}{\cos\tau} \right)^{-\Delta}
 \end{align*}
 with the support to be the region on the boundary inside the bulk lightcone. A simple example is $dS_{1+1}$, for which the support for the smearing function is
 \begin{equation}
   |\rho-\rho'|<\frac{\pi}{2}-\tau
   \label{}
 \end{equation}
 on $\mathcal{I}^+$, where $\rho\in[-\pi,\pi]$ is the spatial coordinate of $dS_{1+1}$.

\begin{figure}[h]
   \centering
   \includegraphics[width=15cm]{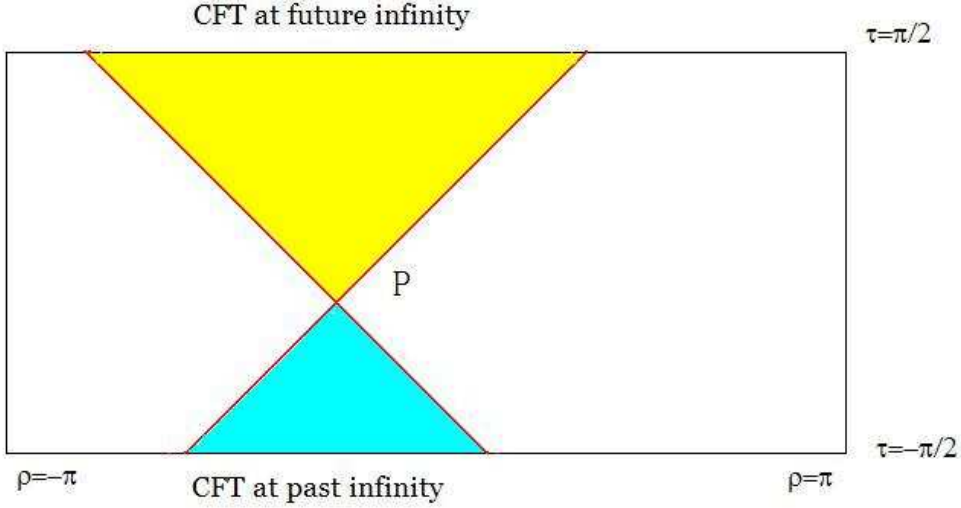}
   \caption{Local Operator In the Global 1+1 dimensional de Sitter space}
   \label{global}
 \end{figure}

 For the smearing functions evolving operators from the past boundary $\mathcal{I}^-$, we have

 \begin{align*}
   & K_+^{\mathcal{I}^-}(\tau,\Omega|\Omega')\sim\left( \frac{\cos(\Omega-\Omega')+\sin\tau}{\cos\tau} \right)^{\Delta-d}\\
   & K_-^{\mathcal{I}^-}(\tau,\Omega|\Omega')\sim\left( \frac{\cos(\Omega-\Omega')+\sin\tau}{\cos\tau} \right)^{-\Delta}
 \end{align*}
 and the support for the case of $dS_{1+1}$, is
 \begin{equation}
   |\rho-\rho'|<\tau+\frac{\pi}{2}
   \label{}
 \end{equation}
 The supports for the $\mathcal{I}^+$ and $\mathcal{I}^-$ smearing functions are each defined within a single lightcone originating from the bulk point and extending to both past and future, as shown in figure (\ref{global}).
 The smearing prescription for a dimension $\Delta$ operator which reduces to the flat de Sitter space smearing function is
 \begin{equation}
 \boxed{  \begin{split}
   \Phi(\tau,\Omega)&=\frac{2^{\Delta-d}\Gamma(\Delta-\frac{d}{2}+1)}{\pi^{\frac{d}{2}}\Gamma(\Delta-d+1)}\int d\Omega'\left( \frac{\cos(\Omega-\Omega')\mp\sin\tau}{\cos\tau} \right)^{\Delta-d}\mathcal{O}_+(\Omega')\\
   &+\frac{2^{-\Delta}\Gamma(\frac{d}{2}-\Delta+1)}{\pi^{\frac{d}{2}}\Gamma(1-\Delta)}\int d\Omega'\left( \frac{\cos(\Omega-\Omega')\mp\sin\tau}{\cos\tau} \right)^{-\Delta}\mathcal{O}_-(\Omega')
 \end{split}}
   \label{}
 \end{equation}
 respectively for $\mathcal{I}^\pm$ smearing, with integration inside the bulk lightcone region. This will reproduce the Wightman function in the Euclidean vacuum, expressed in global coordinates.

 One can construct a bulk operator from $\mathcal{I}^-$ and push it to $\mathcal{I}^+$, in which limit
 \begin{equation}
   \begin{split}
   \Phi(\tau\rightarrow\frac{\pi}{2},\Omega)&\rightarrow\left( \cos\tau \right)^{d-\Delta}\frac{2^{\Delta-d}\Gamma(\Delta-\frac{d}{2}+1)}{\pi^{\frac{d}{2}}\Gamma(\Delta-d+1)}\int d\Omega'\left(\cos(\Omega-\Omega')+1 \right)^{\Delta-d}\mathcal{O}_+(\Omega')\\
   &+\left( \cos\tau \right)^\Delta\frac{2^{-\Delta}\Gamma(\frac{d}{2}-\Delta+1)}{\pi^{\frac{d}{2}}\Gamma(1-\Delta)}\int d\Omega'\left( \cos(\Omega-\Omega')+1 \right)^{-\Delta}\mathcal{O}_-(\Omega')
 \end{split}
 \end{equation}
 This has precisely the form we would expect as the boundary limit of a local bulk operator. But now with the ``$-$'' component expressed by $\mathcal{O}_+$ on $\mathcal{I}^-$ and vice versa:
 \begin{equation}
   \begin{split}
   & \mathcal{O}_+(\Omega,\mathcal{I}^+)=\frac{2^{-\Delta}\Gamma(\frac{d}{2}-\Delta+1)}{\pi^{\frac{d}{2}}\Gamma(1-\Delta)}\int d\Omega'\left( \cos(\Omega-\Omega')+1 \right)^{-\Delta}\mathcal{O}_-(\Omega',\mathcal{I}^-)\\
   & \mathcal{O}_-(\Omega,\mathcal{I}^+)=\frac{2^{\Delta-d}\Gamma(\Delta-\frac{d}{2}+1)}{\pi^{\frac{d}{2}}\Gamma(\Delta-d+1)}\int d\Omega'\left( \cos(\Omega-\Omega')+1 \right)^{\Delta-d}\mathcal{O}_+(\Omega',\mathcal{I}^-)\\
 \end{split}
 \end{equation}
 Here the integration is over the whole past boundary---the past lightcone covers the whole $d$--sphere. The equations above can be regarded as a Bogoliubov transformation in coordinate space, as well as an operator dictionary relating two copies of CFT on $\mathcal{I}^\pm$

 The form of this boundary--boundary map can be better formulated by expressing operators not at angular position $\Omega$, but its antipodal point $\tilde{\Omega}$ instead, since $\cos(\tilde{\Omega}-\Omega')=-\cos(\Omega-\Omega')$  we have
 \begin{equation}
 \boxed{  \begin{split}
   & \mathcal{O}_+(\tilde{\Omega},\mathcal{I}^+)=\alpha_{\Delta,d}\int d\Omega'\langle\mathcal{O}_+(\Omega)\mathcal{O}_+(\Omega')\rangle\mathcal{O}_-(\Omega',\mathcal{I}^-)\\
   & \mathcal{O}_-(\tilde{\Omega},\mathcal{I}^+)=\beta_{\Delta,d}\int d\Omega'\langle\mathcal{O}_-(\Omega)\mathcal{O}_-(\Omega')\rangle\mathcal{O}_+(\Omega',\mathcal{I}^-)
 \end{split}}
 \label{shadow}
 \end{equation}
 with $\alpha$ and $\beta$ being some coefficients that depend on $d$ and $\Delta$, where in global coordinates the CFT two--point function with dimension $\Delta$ is proportional to $\left(\frac{1}{\sin^2\left( \frac{\Omega-\Omega'}{2} \right)}\right)^\Delta$

 The operator relations above between the CFT's at $\mathcal{I}^\pm$ are the dictionary relating two equivalent holographic descriptions of the same bulk. With the state--operator map one can regard them as the transformations relating different bases for a quantum state. Here the collection of operators at either copy of CFT are a complete description for de Sitter space in its Euclidean vacuum state. The total Hilbert space for two CFTs on both boundaries forms a redundant description, with the constraints above necessary to relate half of the space to the other half.

 An interesting way of looking at these operator relations is provided by techniques developed for calculating conformal blocks, in which the relations above Eq. (\ref{shadow}) are actually the definition of shadow operators\footnote{I thank Daliang Li for pointing this out.} \cite{FGPG}. For a given primary operator $\mathcal{O}(x)$ with dimension $\Delta$ in a CFT, its shadow operator $\tilde{O}(x)$ is a non--local operator with dimension $d-\Delta$, which shares the same conformal Casimir with $\mathcal{O}$. To get the part involving a certain primary operator $\mathcal{O}(x)$ in the conformal block decomposition of a CFT four--point function $\langle\varphi_1(x_1)\varphi_2(x_2)\varphi_3(x_3)\varphi_4(x_4)\rangle$, one can insert a projection operator with dimension zero which is defined by both $\mathcal{O}$ and its shadow $\tilde{O}$  \cite{FGPG}\cite{DSD}. Then
 \begin{equation}
   \int d^dx \langle\varphi_1(x_1)\varphi_2(x_2)\mathcal{O}(x)\rangle\langle\tilde{O}(x)\varphi_3(x_3)\varphi_4(x_4)\rangle
   \label{}
 \end{equation}
 gives the conformal block for exchanging the $\mathcal{O}$ operator after projecting out the shadow blocks. In a CFT, for a local primary operator one can always construct its shadow. The shadow operators are non--local operators in the CFT, but they transform like local primary operators under conformal transformations. An explicit relation between a primary operator and its shadow is given in \cite{FKKLPS}:
 \begin{equation}
   \tilde{O}(x)=\int d^dy\frac{1}{\left( x-y \right)^{2\left( d-\Delta \right)}}\mathcal{O}(y)\propto\int d^dy D_{d-\Delta}(x-y)\mathcal{O}(y)
   \label{}
 \end{equation}
 here $D_{d-\Delta}(x-y)$ is the two--point function of a primary operator with dimension $d-\Delta$.

 One can immediately notice that this is a generalization of Eq. (\ref{shadow}), and Eq. (\ref{shadow}) gives a physical interpretation of shadow operators in the special case of Euclidean CFTs on spheres. The shadow for an operator $\mathcal{O}_+$ defined at one of the boundaries of dS, is a local operator $\mathcal{O}_-$ in the CFT defined at the other boundary, with an antipodal map on the sphere. Thus instead of phrasing the construction of de Sitter local operators in terms of two sets of CFT local operators defined at the same boundary, we can also phrase it as a construction with a single copy of operators defined at both boundaries---we use the operator $\mathcal{O}$ defined on one of the boundaries and use its shadow $\tilde{O}$ defined on the other. This also ensures that we are always able to come up with the operators required for the construction: we are always able to construct the shadow operator $\tilde{O}$ from an operator $\mathcal{O}$. Even in the flat slicing, with only a single boundary, the shadow operator for an operator $\mathcal{O}_+$ in the same CFT fits in the properties we need for the corresponding $\mathcal{O}_-$. Thus in the flat patch, the construction can be made by a local operator $\mathcal{O}_+$ and its non--local shadow $\mathcal{O}_-=\int D_-\mathcal{O}_+$. Schematically we have
 \begin{equation}
   \Phi(\eta,\textbf{x})=A_{d,\Delta}\int d^3x' K_{+}\left( \eta,\textbf{x}=0|\textbf{x}' \right) \mathcal{O}_+(\textbf{x}+\textbf{x}')+B_{d,\Delta}\int d^3x'd^3y K_-\left( \eta,\textbf{x}=0|\textbf{x}' \right)D_-\left(\textbf{x}+\textbf{x}'-\textbf{y}  \right)\mathcal{O}_+(\textbf{y})
   \label{}
 \end{equation}

 The second part is from the shadow operator. After integrating over $\textbf{x}'$ inside the lightcone, we have a contribution proportional to
 \begin{equation}
   \int d^dy\left( \frac{\eta^2-\textbf{y}^2}{\eta} \right)^{\Delta-d}\mathcal{O}_+(\textbf{x}+\textbf{y})
   \label{}
 \end{equation}
 The integrand is the same as the contribution from the operator $\mathcal{O}_+$, which is the first term in (\ref{smear}) but the support for the integration is non--compact. The non--compact support is from the definition of the shadow operator, which is the price for expressing a de Sitter local operator with a single CFT operator. In this representation, a bulk operator which is close to the boundary maps to a highly non--local operator: a mixture of a local boundary operator and its shadow.

\subsection{Comments on Gauge Fields in de Sitter Space}
\label{subsec:gauge}
 In the discussion above the attention was on scalar operators in de Sitter space. Here we make some comments on the construction for local fields with integer spins.  We focus on gauge fields propagating in de Sitter space. In \cite{AHS} Vasiliev theory \cite{Va} in de Sitter space is proposed as a higher spin realization of dS/CFT correspondence, with the dual conformal field theory being an $Sp(N)$ model with anti--commuting scalars. In this theory there are infinitely many higher spin conserved currents that are bilinear operators constructed from the scalar multiplet:

 \begin{equation}
   \mathcal{O}_{i_1\dots i_s}=\Omega_{ab}\chi^a\partial_{( i_1}\dots\partial_{i_s)}\chi^b
   \label{}
 \end{equation}
 where $\Omega_{ab}$ is symplectic tensor.

 In \cite{Heem}\cite{KLRS} the holographic constructions for a massless vector field and a graviton field in anti--de Sitter space are established and are generalized to gauge field $\Phi_{M_1\dots M_s}$ with generic integer spin in \cite{XX}.  With the choice of holographic gauge 
 \begin{equation}
   \Phi_{z\dots z}=\Phi_{\mu_1 z\dots z}=\dots=\Phi_{\mu_1\mu_2\dots \mu_{s-1}z}=0
   \label{}
 \end{equation}
 In the Poincar{\'e} patch it is shown that for a generic gauge field with spin $s>1$,

\begin{equation}
  \boxed{ \Phi_{\mu_1\dots\mu_s}=\frac{\Gamma\left( s+\frac{d}{2}-1 \right)}{\pi^{\frac{d}{2}}\Gamma\left( s-1\right)}\frac{1}{z^s}\int_{t'^2+|\textbf{y}'|^2<z^2}dt'd^{d-1}y'\left( \frac{z^2-t'^2-|\textbf{y}'|^2}{z} \right)^{s-2}\mathcal{O}_{\mu_1\dots\mu_s}(t+t',\textbf{x}+i\textbf{y}')}
  \label{}
\end{equation}
where $\mu_i$ are $d$--dimensional indices. The operator $\mathcal{O}_{\mu_1\dots\mu_s}$ is a symmetric traceless conserved current on the AdS boundary. In $d-$dimensional CFT such an operator has dimension $\Delta=s+d-2$. Thus the twist $\Delta-d$ is always $ s-2$, as indicated by the smearing function above.

It is shown in \cite{XX} that it is very convenient to convert $\Phi_{\mu_1\dots\mu_s}$ to a scalar muitiplet with vierbeins in AdS
\begin{equation}
  e_a^{~ ~\mu}=z\delta_a^{~ ~\mu}
  \label{}
\end{equation}
Let us define
\begin{equation}
  Y_{a_1\dots a_s}\equiv e_{a_1}^{~ ~\mu_1}\dots e_{a_s}^{~ ~\mu_s}\Phi_{\mu_1\dots \mu_s}=z^s\Phi_{a_1\dots a_s}
  \label{}
\end{equation}
Here $\Phi_{a_1\dots a_s}$ is written in the sense of components; $\Phi$ itself is still defined as a tensor under diffeomorphism.

One can show that $Y_{a_1\dots a_s}$ obeys a free scalar equation in AdS with mass parameter and scaling dimension
\begin{equation}
  m^2R_{AdS}^2=(s-2)(s+d-2)~,~ \Delta=s+d-2
  \label{}
\end{equation}
and therefore near the boundary
\begin{equation}
  Y\rightarrow z^\Delta\mathcal{O}
  \label{}
\end{equation}
The near boundary behavior of the gauge field is given by:

\begin{equation}
  \Phi_{\mu_1\dots\mu_s}=\frac{1}{z^s}Y_{\mu_1\dots \mu_s}\rightarrow z^{\Delta-s}\mathcal{O}_{\mu_1\dots\mu_s}=z^{d-2}\mathcal{O}_{\mu_1\dots\mu_s}
  \label{}
\end{equation}
Therefore we are able to relate a spin$-s$ bulk gauge field propagating in $AdS_{d+1}$ with a scalar with mass parameter above the Breitenlohner--Freedman bound.

Now we look at the case of gauge fields in de Sitter space.
The Poincar{\'e} patch of AdS can be analytically continued to the flat patch of dS with double analytic continuation:
\begin{equation}
  R_{AdS}^2\rightarrow -R_{dS}^2~,~z\rightarrow \eta~,~x^i_{AdS}\rightarrow ix^i_{ds}
  \label{}
\end{equation}
With the analytic continuation the mass parameter in de Sitter space turns into
\begin{equation}
  m^2R_{dS}^2=-(s-2)(s+d-2)
  \label{}
\end{equation}
The map between scaling dimension and mass parameter in de Sitter space is
\begin{equation}
  \Delta=\frac{d}{2}+\sqrt{\frac{d^2}{4}-m^2R_{dS}^2}
  \label{}
\end{equation}
and thus gives real dimensions
\begin{equation}
  \Delta=s+d-2
  \label{}
\end{equation}
We see that for scalars the scaling dimensions can in general be imaginary, but that for conserved currents, the dimensions are still real integers.

Therefore for a spin$-s>1$ gauge field in de Sitter space, the construction is equivalent to the construction for a massless scalar ($s=2$) or tachyons ($s>2$).
For spin$-1$, the mass of the scalar is $m^2=d-1$ which is positive, but in general for a $d-$dimensional space\footnote{With the exception of $d=2$, which satuates the bound, which suggests that maybe the bulk Chern-Simons field has something special in the story here.}, $d-1<\frac{d^2}{4}$ and thus does not satisfy $m^2>\left( \frac{d}{2} \right)^2$. Therefore we see that a local observable for a gauge field in de Sitter space behaves quite differently from the heavy scalar operators we have constructed. These gauge field operators have real dimensions and when approaching the boundaries, components with $\left( \sigma\eta' \right)^\Delta$ and $\left( \sigma\eta' \right)^{d-\Delta}$ fall at different rates, and have no oscillatory behaviors.

One can see this difference explicitly when directly extending the construction above to gauge fields. One might naively expect that for gauge fields the construction involves two sets of single$-$trace operators with dimensions $\Delta$ and $d-\Delta$ and gives the construction equation below:
\begin{align*}
  \Phi_{i_1\dots i_s}&=\frac{\Gamma\left( s+\frac{d}{2}-1 \right)}{\pi^{\frac{d}{2}}\Gamma\left( s-1\right)}\frac{1}{\eta^s}\int_{\textbf{x}'^2<\eta^2}d^dx'\left( \frac{\eta^2-\textbf{x}'^2}{\eta} \right)^{s-2}\mathcal{O}^{(+)}_{i_1\dots i_s}(\textbf{x}+\textbf{x}')\\
  &+\frac{\Gamma(3-s-\frac{d}{2})}{\pi^{\frac{d}{2}}\Gamma(3-s-d)}\frac{1}{\eta^s}\int_{\textbf{x}'^2<\eta^2}d^dx'\left( \frac{\eta^2-\textbf{x}'^2}{\eta} \right)^{2-s-d}\mathcal{O}^{(-)}_{i_1\dots i_s}(\textbf{x}+\textbf{x}')
\end{align*}
which is obtained by simply substituting the dimension $\Delta=s+d-2$ into the scalar expression and identifying $Y_{i_1\ldots i_s}=\eta^s \Phi_{i_1\ldots i_s}$as a bulk scalar. 

One would then notice that this proposed solution has problems: The first term has a diverging denominator when $s=1$ and the second term has a diverging denominator when $d+s$ is an integer larger than two. Therefore, for all the cases of interest, $s$ and $d$ taking values on positive integers, the construction equation above is not well--defined.

The root of the problem is the starting point of the construction for these fields. They are constructed by demanding $\eta^s\Phi_{i_1\ldots i_s}$ as scalars recover the Wightman function for Euclidean vacuum, Eq (\ref{wight}). Here in general $\eta^s\Phi_{i_1\dots i_s}$ are scalars with mass parameters that go below  $\left( \frac{d}{2} \right)^2$. Also for the case of interest, $s$ and $d$ are integers, and so the Wightman function is ill--defined for $s>1$ due to the factor involving $\Gamma(1-\Delta=3-s-d)$, and for spin $s=1$ it doesn't exhibit the nice property of spliting into two parts with fall--off behaviors $\eta^\Delta$ and $\eta^{d-\Delta}$ because the hypergeometric function behaves in a different way when its arguments are integers. 

Interestingly enough, these operators are exactly the ones relevant for the proposal of duality between the $Sp(N)$ model in 3 dimensions and $dS_4$ \cite{AHS}. There we have currents $J_{i_1\dots i_s}=\Omega_{ab}\chi^a\partial_{(i_1}\dots\partial_{i_s)}\chi^b$ with dimension $s+1$, which correspond to mass parameter $m^2=(2-s)(s+1)$, due to these values for the mass parameter, for these fields the approach above fails and we no longer have a nice picture of a bulk operator being constructed from a pair of CFT operators, and recovering bulk Wightman function in a de Sitter invariant vacuum.

Despite the remarks above, we can still get a bulk operator that is dual to a boundary current, in the following sense. Given a boundary spin$-s$ current there is an operator in the bulk that matches it smoothly when approaching the boundary.  To get such an operator, one can just keep the first part of the expression which approaches the boundary higher spin current:
\begin{equation}
  \boxed{  \Phi^{(+)}_{i_1\dots i_s}=\frac{\Gamma\left( s+\frac{d}{2}-1 \right)}{\pi^{\frac{d}{2}}\Gamma\left( s-1\right)}\frac{1}{\eta^s}\int_{\textbf{x}'^2<\eta^2}d^dx'\left( \frac{\eta^2-\textbf{x}'^2}{\eta} \right)^{s-2}\mathcal{O}^{(+)}_{i_1\dots i_s}(\textbf{x}+\textbf{x}')}
  \label{}
\end{equation}

Here we still have the singularity for a vector field $s=1$ from the coefficient. A more careful treatment following \cite{KLRS}\cite{XX} gives the expression
\begin{equation}
  \boxed{  A_i^{(+)}=\frac{1}{V(S^{d-1})}\frac{1}{\eta}\int_{|\textbf{x}'|=\eta}d^{d-1}x'\mathcal{J}_i^{(+)}\left( \textbf{x}+\textbf{x}' \right)}
  \label{}
\end{equation}
Here the integration is over the intersection of the bulk lightcone and the boundary which is a sphere, and $V\left( S^{d-1} \right)=\frac{2\pi^{\frac{d}{2}}}{\Gamma\left( \frac{d}{2} \right)}$ is the surface area of a ($d-1$)--sphere.
This is just the analytically continued version of the construction in anti--de Sitter space. Here we are actually imposing Dirichlet type boundary condition at $\mathcal{I}^\pm$ by demanding that $\Phi_{\mu_1\dots \mu_s}\rightarrow \eta^{\Delta-s}\mathcal{O}_{\mu_1\dots\mu_s}$. As has been discussed in \cite{ANS}, such Dirichlet boundary conditions are acausal---they force the radiation that hits the future boundary to reflect back into the past. As was discussed in the section 3, the boundary conditions kill the bulk positive or negative mode, and thus spoil microcausality. The correlation functions computed with such operators are the ones we can obtain by analytically continuing the AdS correlation functions, and thus do not correspond to any de Sitter invariant vacuum. They are not the operators for computing the correlation functions if one would like to look at the cosmology in de Sitter space.

Looking back at gauge fields in AdS space one notices that the situation is much simpler: the mass parameter for the scalar which corresponds to a spin$-s$ gauge field is always within the Breitenlohner--Freedman bound $m^2>-\frac{d^2}{4}$. The constructions can be carried out in a standard way \cite{XX}. 

\subsection{Implementation in the Embedding Formalism}

We can ask if it is possible to recast the constructions above in the language of the embedding formalism, which was developed in \cite{Dirac}\cite{MS}\cite{Wei}, and recently generalized for superconformal field theories in \cite{GSS}\cite{GKLS}. In this language, the conformal group in $d$ dimensions, which is $SO(d,2)$, is realized as the Lorentz group in a $d+2$ dimensional Minkowski space with two time directions. The transformations on the $d+2$ coordinates are linear, and the conformally invariant quantities in $d$ dimensions can be built as Lorentz invariant quantities in $d+2$ dimensional embedding space, which shows the conformal invariance manifestly.

The set--up of the embedding formalism starts with a $d+2$ dimensional Minkowski space with two time--like directions:
\begin{equation}
  ds^2=\eta_{IJ}dX^IdX^J=-dX^+dX^-+\delta_{mn}dX^mdX^n
  \label{}
\end{equation}
where the indices $I,J$ run over $d+2$ coordinates and $m,n$ run over $d$ coordinates.

The anti--de Sitter space and de Sitter space are realized as the hypersurfaces defined by:
\begin{equation}
  X_{AdS}\cdot X_{AdS}=-R_{AdS}^2~,~X_{dS}\cdot X_{dS}=R_{dS}^2
  \label{}
\end{equation}
At large $X$ both the hypersurfaces for de Sitter and for anti--de Sitter space approach the ($d+2$)--dimensional embedding space lightcone, which is 
\begin{equation}
  X\cdot X=0
  \label{}
\end{equation}
We can define a $d$--dimensional Minkowski space by turning the ($d+2$)--dimensional embedding space lightcone into a projective space, denoting the points in the embedding space on the lightcone as $P^I$. We then demand that $P^I$ satisfy:
\begin{equation}
  P\cdot P=0~,~ P^I\sim \lambda P^I
  \label{}
\end{equation}
Here we identify the points on the embedding space lightcone that are on the same ray from the origin, thus forming a $d$ dimensional space.

One can parametrize the embedding space in the following ways. To recover the $d$--dimensional Minkowski space, we define coordinates on the projective lightcone as
\begin{equation}
  P^I=\left( 1, y^2, y^\mu \right)
  \label{}
\end{equation}
where $y^\mu$ are $d$--dimensional coordinates. The distance between two points on the projective lightcone is then $-2P_1\cdot P_2=(y_1-y_2)^2$. We see that it recovers the distance between two points in the Minkowski spacetime.

For $AdS_{d+1}$ one can define:
\begin{equation}
  X_{AdS}^I=\frac{1}{z}\left( 1, z^2+x^2, x^\mu \right)
  \label{}
\end{equation}
We then have the distance between two points in AdS space as
\begin{equation}
  -2X_{AdS,1}\cdot X_{AdS,2}=\frac{z_1^2+z_2^2+(x_1-x_2)^2}{z_1z_2}
  \label{}
\end{equation}
which is proportional to the distance $\sigma$ we used in the sections above.

Also the regularized distance between a boundary and a bulk point is
\begin{equation}
  -2P\cdot X=\frac{z^2+(x-y)^2}{z}
  \label{}
\end{equation}
Now we see that in anti--de Sitter space, with the embedding coordinates, we can write down the smearing function in a very simple and manifestly AdS--invariant way:
\begin{equation}
\boxed{  \Phi(X)=A_{\Delta,d}\int_{\partial AdS}dP\left( -2P\cdot X \right)^{\Delta-d}\Theta\left( -P\cdot X \right)\mathcal{O}(P)}
  \label{}
\end{equation}
Here we integrate over the boundary points denoted by $P$, to get a certain bulk operator sitting at point $X$ in the embedding coordinates. The domain of integration is over the region with $P\cdot X<0$, where the boundary points are spacelike separated from the bulk point in the $z$ direction.

In the construction of causal three--point functions in AdS \cite{KLL}, there is an AdS--invariant cross--ratio which is particularly interesting:
\begin{equation}
  \chi\left( z,x;x_1;x_2 \right)=\frac{\left( z^2+(x-x_1)^2 \right)\left( z^2+(x-x_2)^2 \right)}{z^2\left( x_2-x_1 \right)^2}
  \label{}
\end{equation}
This cross--ratio is an AdS--invariant quantity built from a single bulk point and two boundary points. It turns out that in AdS space the towers of multi--trace operators to be added into the smearing prescription for recovering bulk microcausality at the level of $N^{-1}$ are organized by powers of this cross--ratio \cite{KLL}. In the embedding formalism, this cross--ratio is simple:
\begin{equation}
  \chi\left( z,x;x_1;x_2 \right)=4\frac{\left( P_1\cdot X \right)\left( P_2\cdot X \right)}{P_1\cdot P_2}
  \label{}
\end{equation}

where $X$ denotes the bulk point and $P_i$ for the boundary points.

We can also describe the smearing prescription in de Sitter space with the embedding formalism, starting from embedding de Sitter space into the higher dimensional Minkowski space:
\begin{equation}
  X_{dS}^I=\frac{1}{\eta}\left( 1, -\eta^2+\textbf{x}^2, x^i \right)
  \label{}
\end{equation}
We then have the distance between two points in de Sitter space:
\begin{equation}
  -2X_1\cdot X_2=\frac{-\eta_1^2-\eta_2^2+(\textbf{x}_1-\textbf{x}_2)^2}{\eta_1\eta_2}
  \label{}
\end{equation}
and the regularized distance between a bulk point and a boundary point is:
\begin{equation}
  -2P\cdot X=\frac{-\eta^2+(\textbf{x}-\textbf{y})^2}{\eta}
  \label{}
\end{equation}

Therefore for a scalar with $m^2>\left( \frac{d}{2} \right)^2$ we have the smearing prescription in the embedding space:
\begin{equation}
  \boxed{  \Phi(X)=A_{\Delta,d}\int_{\partial dS} dP\left( 2P\cdot X \right)^{\Delta-d}\Theta\left( P\cdot X \right)\mathcal{O}_+(P)+B_{\Delta,d}\int_{\partial dS} dP\left( 2P\cdot X \right)^{-\Delta}\Theta\left( P\cdot X \right)\mathcal{O}_-(P)}
  \label{}
\end{equation}

The dS--invariant cross--ratio
\begin{equation}
  \chi\left( \eta,\textbf{x};\textbf{x}_1;\textbf{x}_2 \right)=4\frac{\left( P_1\cdot X \right)\left( P_2\cdot X \right)}{P_1\cdot P_2}=\frac{\left( -\eta^2+(\textbf{x}-\textbf{x}_1)^2 \right)\left( -\eta^2+(\textbf{x}-\textbf{x}_2)^2 \right)}{\eta^2(\textbf{x}_2-\textbf{x}_1)^2}
  \label{}
\end{equation}
could be useful when one considers microcausality for three--point functions in de Sitter space.

It could be interesting to perform the construction for gauge fields in the embedding space, which could potentially make the AdS invariance manifest in the construction. In \cite{KLRS}\cite{XX} the construction is done in AdS space by imposing the holographic gauge. The construction is not done in a manifestly AdS covariant way, therefore one has to check the AdS covariance of the constructions afterwards. The embedding formalism could be helpful in that direction. Here we didn't derive the construction starting from the embedding formalism, but rather just wrote down the final results in the embedding space. It also could be interesting if we can solve the Cauchy problem or sum over the modes starting with the embedding formalism and derive the equations above. 

%%%%%%%%%%%%%%%%%%%%%%%%%%%%%%%%%
\section{Discussion and Outlook}%
%%%%%%%%%%%%%%%%%%%%%%%%%%%%%%%%%

In this paper we made progress in the construction of de Sitter bulk operators in terms of non--local boundary operators. For heavy scalars with $m^2>\left( \frac{d}{2} \right)^2$ the construction recovers the bulk Wightman function in the Euclidean vacuum state. Here we should emphasize that the construction is state--dependent: for different vacuum states in de Sitter space, we have different de Sitter and CFT correlation functions, and thus different construction prescriptions relating bulk observables and boundary operators. In de Sitter space there are other invariant vacua apart from the Euclidean vacuum, especially a one--parameter family known as ``$\alpha$--vacua''. How should we construct operators so that we recover the bulk correlators in these vacua? In \cite{BMS} the dependence of the bulk Wightman function on $\alpha$ was worked out in the limit that the points approach the boundaries. It would be interesting to see the $\alpha$ dependence of bulk operators.
 
 We have performed the construction at the level of two point function, which is at the order of $N^0$ in the large--N expansion. To go beyond this one would like to think about three--point functions with each of the bulk operators corresponding to two boundary operators, there can be more subtleties than in AdS case. For instance, the three--point function of scalars in de Sitter space, with one scalar deep inside de Sitter and two others close to the boundary, is schematically
 \begin{equation*}
   \langle\Phi(\eta_1,\textbf{x}_1)\Phi(\eta_2\sim 0,\textbf{x}_2)\Phi(\eta_3\sim 0, \textbf{x}_3)\rangle\sim(\eta_2\eta_3)^\Delta\int K_+\langle\mathcal{O}_+\mathcal{O}_+\mathcal{O}_+\rangle+\eta_2^\Delta\eta_3^{d-\Delta}\int K_+\langle\mathcal{O}_+\mathcal{O}_+\mathcal{O}_-\rangle+\dots
   \label{}
 \end{equation*}
which involves all the boundary three--point functions that are constructed from $\mathcal{O}_\pm$.  As in the AdS case, one would expect that the bulk lightcone singularities will show up in each of these terms, thus breaking microcausality. The hope is to include towers of multi--trace operators into the construction and recover microcausality. We should emphasize that to check microcausality at the level of three--point funtions one can no longer just start with three--point functions like $<\Phi(\eta,\textbf{x})\mathcal{O}_\pm(\textbf{x}_1)\mathcal{O}_\pm(\textbf{x}_2)>$, i.e. the correlation function between a bulk field and boundary fields with a single dimension, because now a boundary operator with a single scaling dimension doesn't match with the near--boundary limit of any local bulk field. This is different from the construction in AdS, in which a local bulk field, when approaching the boundary, matches smoothly onto a boundary operator with a single dimension. This is a nice property of AdS/CFT which doesn't hold when one goes into de Sitter space. 

 One can also consider bulk operators in de Sitter space with other slicings, such as what we have done for the flat and the global slicings. The static patch would be a very interesting case. Once we go into the static patch, we have no asymptotic boundaries anymore---they are behind the horizon, so one can ask in that case what data should be used to construct the local operators. Possibly the approaches people take for the problem of constructing local observables behind the horizon of an eternal black hole in AdS could shed some light on it. There though the interior is separated from the boundary by a horizon, still the construction from the boundary is shown to be possible with extra degrees of freedom involved---either from the CFT's thermofield double \cite{HKLL2} or from a fine--grained sector in CFT \cite{PR}. In de Sitter space, it is not clear if a similar construction will work. There are proposals for static patch holography such as dS/dS\cite{dSdS}\cite{micro} and static patch solipsism\cite{soli}, in which the dual theory to the static patch lives on the central slice of dS slicing and the central worldline respectively. It would be interesting to explore whether and how the discussion of operator dictionaries and bulk locality can be extended to these proposals. A related question is how to understand local fields when bubble nucleation is considered. We have already mentioned the possible subtleties which may potentially falsify the existence of dS/CFT correspondence. One of them is the metastability of de Sitter space. As eternal inflation populates the landscape, what appears in the asymptotic future may be a fractal of FRW universes in nucleated bubbles, instead of a flat Euclidean space on which we can define a Euclidean CFT. Since our construction is state--dependent, it refers to a certain background spacetime. Thus the nucleation of bubbles could potentially modify the construction prescription dramatically. However from the point of view of a FRW universe as a semi--classical background, the notion of microcausality should be well--defined. A proposal for a holographic description of an FRW universe in the Coleman--de Luccia bubble is described in \cite{FSSY}, aiming to giving a holographic description to eternal inflation. Thus one may think about how to formulate local bulk physics in terms of the dual data in such a proposal. 
 
 All the constructions that we presented above are about empty de Sitter space in certain vacuum states. One can also think about black holes in de Sitter space. The construction of local operators behind the event horizon of an eternal black hole in anti--de Sitter space has been performed by several authors, such as in \cite{HKLL2}, where the construction is established for the BTZ black hole in $AdS_{2+1}$ as a special case, and a local operator inside a black hole is shown to involve operators in both a conformal field theory and its thermo--field double. In \cite{PR} the construction is generalized to Schwartzchild black holes in AdS of generic dimensions. The construction data one uses are from a single copy of CFT, established with the help of ``mirror operators''. It would be interesting to think about black holes in de Sitter space, how the structure of a CFT with its thermo--field double is realized in de Sitter black hole, and how observables behind the black hole horizon can be constructed.

 Finally, one can ask if we can understand bulk locality using the data on a null surface. Just from the point of view of an initial value problem, there are types of well--defined null initial value problems such as the Bondi problem\cite{Bondi} and Sachs' double null problem\cite{Sachs}. However it is not clear if this kind of initial data can be interpreted as a kind of field theory in lower dimensions. This is similar to the hope of describing spacetime with vanishing cosmological constant with some degrees of freedom out at null infinity \cite{QdS}. Recently the series of papers on asymptotic symmetries in asymptotically Minkowski space\cite{Stro}\cite{Stro2} may suggest such a theory. However, determining what kind of degrees of freedom could live at null boundary and what kind of dynamics could be candidates is still speculative.

 %%%%%%%%%%%%%%%%%%%%%%%%%%%%
 \section*{Acknowledgements}%
 %%%%%%%%%%%%%%%%%%%%%%%%%%%%

 I am grateful to Dionysios Anninos, Frederik Denef, Bart Horn, Daniel Kabat, Daliang Li, Debajyoti Sarkar for inspiring discussions and comments on the draft. Special thank to Bart Horn and Erick Weinberg for detailed review of the draft and giving valuable suggestions. The work was supported in part by the U.S. Department of Energy grant DE-FG02-92ER40699.

 \begin{appendices} 
   \section{Integration}
Here we evaluate the integral
 \begin{equation}
   f(\alpha,\beta)=\int_{|\textbf{x}'|<\eta}d^dx'\left( \frac{\eta^2-\textbf{x}'^2}{\eta} \right)^\alpha\frac{1}{\left( \textbf{x}+\textbf{x}' \right)^{2\beta}}
   \label{}
 \end{equation}
 For convenience we make the choice $x^1=|\textbf{x}|\equiv R$, $x^2=\dots=x^d=0$, thus we have
 \begin{equation}
   f(\alpha,\beta)=Vol(S^{d-2})\int^\eta_0 dr r^{d-1}\left( \frac{\eta^2-r^2}{\eta} \right)^\alpha\int^\pi_0\frac{\sin^{d-2}\theta d\theta}{\left( R^2+2Rr\cos\theta+r^2 \right)^\beta}
   \label{}
 \end{equation}
 where
 \begin{equation}
   Vol(S^{d-2})=\frac{2\pi^{\frac{d-1}{2}}}{\Gamma(\frac{d-1}{2})}
   \label{}
 \end{equation}
 Using the formulae:

\begin{align*}
  & \int^\pi_0\frac{\sin^{2\mu-1}\theta}{\left( 1+2a\cos\theta+a^2 \right)^\nu}d\theta=\frac{\Gamma(\mu)\Gamma(\frac{1}{2})}{\Gamma(\mu+\frac{1}{2})}F\left( \nu,\nu-\mu+\frac{1}{2},\mu+\frac{1}{2},a^2 \right)\\
  & \int^1_0 (1-x)^{\mu-1}x^{\gamma-1}F(\alpha,\beta,\gamma,ax)dx=\frac{\Gamma(\mu)\Gamma(\gamma)}{\Gamma(\mu+\gamma)}F(\alpha,\beta,\gamma+\mu,a)
\end{align*}
we then have

 \begin{equation}
   f\left( \alpha,\beta \right)=\frac{\pi^{\frac{d}{2}}\Gamma(\alpha+1)}{\Gamma(\alpha+\frac{d}{2}+1)}\frac{\eta^{\alpha+d}}{|\textbf{x}|^{2\beta}}F\left( \beta,\beta-\frac{d}{2}+1,\alpha+\frac{d}{2}+1,\frac{\eta^2}{\textbf{x}^2} \right)
   \label{}
 \end{equation}
 Also to get the near--boundary two--point Wightman function we need the following property of the hypergeometric function:
 \begin{equation}
   F\left( \alpha,\beta,\beta,z \right)=\left( 1-z \right)^{-\alpha}
   \label{}
 \end{equation}
 \section{Smearing Function from Green's Function}

In de Sitter space, an operator near the past boundary can be expressed by operators in the CFT:
\begin{equation}
  \Phi(\eta\rightarrow 0, \textbf{x})\sim \eta^\Delta\mathcal{O}_++\eta^{d-\Delta}\mathcal{O}_-
  \label{}
\end{equation}
Then if we want to probe deeper into de Sitter space, we need something like
\begin{equation}
  \Phi(\eta,\textbf{x})=\int K_+(\eta,\textbf{x}|\textbf{x}')\mathcal{O}_+(\textbf{x}')+\int K_-(\eta,\textbf{x}|\textbf{x}')\mathcal{O}_-(\textbf{x}')
  \label{}
\end{equation}
Here it is important that we have both components, which means that the correct construction of local operator in de Sitter space is not an analytic continuation from anti de Sitter space; otherwise, we spoil microcausality.

Here the bulk operator is linked to boundary CFT operators with a retarded Green's function defined as 
\begin{equation}
  G_{ret}(x,x')\equiv G_E(x,x')-G_E(x',x)
  \label{}
\end{equation}
with $G_E$ being the Wightman function in Euclidean vacuum.

Here we just need the asymptotic form of $G_{ret}$ in the limit $\eta'\rightarrow 0$:
  
\begin{equation}
  G_{ret}|_{\eta'\rightarrow 0}\sim c_{\Delta,d}(-\sigma-i\epsilon)^{\Delta-d}+c_{\Delta,d}^*(-\sigma-i\epsilon)^{-\Delta}-c.c
  \label{}
\end{equation}
The bulk operator is evaluated by:

\begin{equation}
  \Phi(\eta,\textbf{x})=\int_{|\textbf{x}'|<\eta}d^dx'\left( \frac{1}{\eta'} \right)^{d-1}\left( G_{ret}(\eta,\textbf{x};\eta',\textbf{x}')\partial_{\eta'}\Phi(\eta',\textbf{x}')-\Phi(\eta',\textbf{x}')\partial_{\eta'}G_{ret}(\eta,\textbf{x};\eta',\textbf{x}') \right)
  \label{}
\end{equation}
Taking partial derivative on the retarded Green's function and working in small $\eta'$ limit, we have
\begin{equation}
  \partial_{\eta'}G_{ret}\sim\frac{1}{\eta'}\left( c(\Delta-d)(-\sigma-i\epsilon)^{\Delta-d}-c^*\Delta(-\sigma-i\epsilon)^{-\Delta}+c^*\Delta(-\sigma+i\epsilon)^{-\Delta}-c(\Delta-d)(-\sigma+i\epsilon)^{\Delta-d} \right)
  \label{}
\end{equation}
Therefore for $\Phi_+\sim \eta^\Delta\mathcal{O}_+$ we have
\begin{align*}
  & \left( \frac{1}{\eta'} \right)^{d-1} \left[\Phi_+(\eta',\textbf{x}')\partial_{\eta'}G_{ret}(\eta,\textbf{x}|\eta',\textbf{x}')-G_{ret}(\eta,\textbf{x}|\eta',\textbf{x}')\partial_{\eta'}\Phi_+(\eta',\textbf{x}')\right]\\
  & =\left(\eta'\right)^{\Delta-d}\left[ cd\left( (-\sigma+i\epsilon)^{\Delta-d}-(-\sigma-i\epsilon)^{\Delta-d} \right)+2c^*\Delta\left( (-\sigma+i\epsilon)^{-\Delta}-(-\sigma-i\epsilon)^{-\Delta} \right) \right]\mathcal{O}_+
\end{align*}
Here the factor $\left(\eta'\right)^{\Delta-d}$ cancels with the factor of $\eta'$ with the inverse power from $\sigma^{\Delta-d}$ and gives a well-defined limit when $\eta'\rightarrow 0$, but it doesn't cancel with the factor in $\sigma^{-\Delta}$, leading to a fast oscillation when $\eta'\rightarrow 0$ so the term proportional to $\sigma^{-\Delta}$ vanishes.

For $\Phi_-\sim \eta^{d-\Delta}\mathcal{O}_-$ we have
\begin{align*}
  & \left( \frac{1}{\eta'} \right)^{d-1} \left[\Phi_-(\eta',\textbf{x}')\partial_{\eta'}G_{ret}(\eta,\textbf{x}|\eta',\textbf{x}')-G_{ret}(\eta,\textbf{x}|\eta',\textbf{x}')\partial_{\eta'}\Phi_-(\eta',\textbf{x}')\right]\\
  & =\eta'^{-\Delta}\left[2c(\Delta-d)\left( (-\sigma-i\epsilon)^{\Delta-d}-(-\sigma+i\epsilon)^{\Delta-d} \right)+c^*d\left( (-\sigma+i\epsilon)^{-\Delta}-(-\sigma-i\epsilon)^{-\Delta} \right)\right]\mathcal{O}_-
\end{align*}
Similarly we only have the contribution from the terms proportional to $\sigma^{-\Delta}$

To evaluate the integration kernel, we notice that outside the bulk lightcone $\sigma\propto \eta^2-(\textbf{x}-\textbf{x}')^2<0$ so the $\epsilon$ prescription can be dropped and the integral gives a vanishing result. When we analytically continue the result into the bulk lightcone, the $\epsilon$ prescription will give a phase shift proportional to $Im\left( \Delta-d \right)$ and $Im\left( -\Delta \right)$ respectively. For instance, in
\begin{equation}
  (-\sigma+i\epsilon)^{\Delta-d}
  \label{}
\end{equation}
the cut starts from $\sigma=i\epsilon$ and to analytically continue we go under the branch point and thus get a phase $e^{-i\pi\left( i\sqrt{m^2-\frac{d^2}{4}} \right)}$
and therefore 
\begin{equation}
  (-\sigma+i\epsilon)^{\Delta-d}- (-\sigma-i\epsilon)^{\Delta-d}=-2ie^{i\pi(\Delta-d)}\sin\left( \pi i\left( \Delta-d \right)\right)\sigma^{\Delta-d}
  \label{}
\end{equation}
In this way for $\mathcal{O}_+$ we have a smearing function proportional to $(\sigma\eta')^{\delta-d}$ and for $\mathcal{O}_-$ we have a smearing function proportional to $(\sigma\eta')^{-\Delta}$

\section{ A Property of Hypergeometric Function}
In finding the asymptotic form of the scalar two--point Wightman function, we find the relation below very useful:
\begin{align*}
  F(\alpha,\beta,\gamma;z)&=\frac{\Gamma(\gamma)\Gamma(\beta-\alpha)}{\Gamma(\gamma-\alpha)\Gamma(\beta)}(-z)^{-\alpha}F\left( \alpha,\alpha-\gamma+1,\alpha-\beta+1;\frac{1}{z} \right)\\
 & +\frac{\Gamma(\gamma)\Gamma(\gamma-\alpha-\beta)}{\Gamma(\alpha)\Gamma(\gamma-\beta)}(-z)^{-\beta}F\left( \beta,\beta-\gamma+1,\beta-\alpha+1;\frac{1}{z} \right)
\end{align*}
We see here that the function nicely splits into two parts with behaviors $z^{-\alpha}$ and $z^{-\beta}$ when $z\rightarrow\infty$, and they give the two components with different scaling dimensions in Wightman function.

This relation is true when neither $\alpha-\beta$ nor $\gamma-\alpha-\beta$ is an integer, and is thus applicable to the case when a de Sitter scalar has mass parameter
\begin{equation}
  m^2>\left( \frac{d}{2} \right)^2
  \label{}
\end{equation}
as well as to light particles with non--integer dimensions. However, for gauge fields the dimensions are integers determined by the spin and spatial dimension, so this property is not applicable.

\section{Higher Spin Fields in AdS and dS}

Here we briefly review some results about general integer spin gauge field in $AdS_{d+1}$, following \cite{XX}

Massless gauge fields in AdS are represented by totally symmetric rank--s tensors $\Phi_{M_1\dots M_s}$ satisfying double--tracelessness conditions;
\begin{align}
   \Phi^{MN}_{~ ~ ~ ~MN M_5\dots M_s}=0
  \label{}
\end{align}
The linear equation for a spin--s gauge field on AdS is \cite{Va}
\begin{equation}
  \nabla_N\nabla^N\Phi_{M_1\dots M_s}-s\nabla_N\nabla_{M_1}\Phi^N_{~M_2i\dots M_s}+\frac{1}{2}s(s-1)\nabla_{M_1}\nabla_{M_2}\Phi^N_{~N\dots M_s}-2(s-1)(s+d-2)\Phi_{M_1\dots M_3}=0
  \label{}
\end{equation}
This equation is invariant under the gauge transformation
\begin{equation}
  \Phi_{M_1\dots M_s}\rightarrow \Phi_{M_1\dots M_s}+\nabla_{M_1}\Lambda_{M_2\dots M_s}~,~ \Lambda^N_{~ ~NM_3\dots M_s}=0
  \label{}
\end{equation}
We can choose the holographic gauge in which all the $z$--components of the gauge field vanish \cite{XX}
\begin{equation}
  \Phi_{z\dots z}=\Phi_{\mu_1 z\dots z}=\dots =\Phi_{\mu_1\dots\mu_{s-1}z}=0
  \label{}
\end{equation}

The bulk gauge field is dual to a totally symmetric, traceless, conserved rank--s tensor on the boundary:
\begin{equation}
  \mathcal{O}^\nu_{~\nu \mu_3\dots\mu_s}=0~,~\partial_\nu \mathcal{O}^\nu_{~\mu_3\dots\mu_s}=0
  \label{}
\end{equation}
Therefore, to be consistent we have to set
\begin{equation}
  \Phi^\nu_{~ ~\nu\mu_3\dots\mu_s}=0~,~\partial_\nu\Phi^\nu_{~ ~\mu_3\dots\mu_s}=0
  \label{}
\end{equation}
thus we get:

\begin{equation}
  \left( \partial_z^2+\partial_\alpha\partial^\alpha \right)\Phi_{\mu_1\dots\mu_s}+\frac{2s+1-d}{z}\partial_z\Phi_{\mu_1\dots\mu_s}+\frac{2(s-1)(2-d)}{z^2}\Phi_{\mu_1\dots\mu_s}=0
  \label{}
\end{equation}
 We define
\begin{equation}
  Y_{\mu_1\dots \mu_s}=z^s\Phi_{\mu_1\dots\mu_s}
  \label{}
\end{equation}
as a multiplet of scalars. And the equation for $Y_{\mu_1\dots\mu_s}$ is
\begin{equation}
  \partial_\alpha\partial^\alpha Y_{\mu_1\dots\mu_s}+z^{d-1}\partial_z\left( z^{1-d}\partial_z Y_{\mu_1\dots\mu_s} \right)-\frac{(s-2)(s+d-2)}{z^2}Y_{\mu_1\dots\mu_s}=0
  \label{}
\end{equation}
which is just free scalar equation with mass parameter
\begin{equation}
  m^2=(s-2)(s+d-2)
  \label{}
\end{equation}
corresponding to scaling dimension
\begin{equation}
  \Delta=s+d-2
  \label{}
\end{equation}
The near--boundary behavior of $Y_{\mu_1\dots\mu_s}$ is
\begin{equation}
  Y_{\mu_1\dots\mu_s}\sim z^\Delta\mathcal{O}_{\mu_1\dots\mu_s}
  \label{}
\end{equation}
So one can directly construct the bulk spin--s field:
\begin{equation}
  \Phi_{\mu_1\dots\mu_s}=\frac{\Gamma\left( s+\frac{d}{2}-1 \right)}{\pi^{\frac{d}{2}}\Gamma\left( s-1\right)}\frac{1}{z^s}\int_{t'^2+|\textbf{y}'|^2<z^2}dt'd^{d-1}y'\left( \frac{z^2-t'^2-|\textbf{y}'|^2}{z} \right)^{s-2}\mathcal{O}_{\mu_1\dots\mu_s}(t+t',\textbf{x}+i\textbf{y}')
  \label{}
\end{equation}
for fields with integer spin $s>1$

We see the field behaves like $z^{\Delta-s}=z^{d-2}$ near the boundary.

The reason why $Y_{\mu_1\dots \mu_s}$ turns out to be a scalar is that it is actually the components of the gauge field in a vierbein basis. In AdS we have that
\begin{equation}
  e_a^{~ ~\mu}=z\delta_a^{~ ~\mu}
  \label{}
\end{equation}
and
\begin{equation}
  e_{a_1}^{~ ~\mu_1}\dots e_{a_s}^{~ ~\mu_s}\Phi_{\mu_1\dots\mu_s}\equiv Y_{a_1\dots a_s}
  \label{}
\end{equation}
are scalars because they don't actually carry any spacetime indices---they are defined with respect to a certain vierbein basis at each point in the spacetime.
It is a bit of abuse of the notation not to distinguish $Y_{a_1\dots a_s}$ and $Y_{\mu_1\dots\mu_s}$ but at the end of the day we multiply the inverse vierbeins and recover $\Phi_{\mu_1\dots\mu_s}$ and it does not matter whether we make the vierbeins explicit.

In de Sitter space, the free field equation for $Y_{i_1\dots i_s}$ is obtained by direct analytic continuation:
\begin{equation}
  \ddot{Y}_{i_1\dots i_s}+\frac{1-d}{\eta}\dot{Y}_{i_1\dots i_s}+\left( \frac{(2-s)(s+d-2)}{\eta^2}-\partial_j^2 \right)Y_{i_1\dots i_s}=0
  \label{}
\end{equation}
which matches with the generic form of scalar equations in dS:
\begin{equation}
  \left( \Box-m^2 \right)\phi=0 \rightarrow \ddot{\phi}+\frac{1-d}{\eta}\dot{\phi}+\left(\frac{m^2}{\eta^2}-\partial_j^2  \right)\phi=0
  \label{}
\end{equation}
Thus $Y_{i_1\dots i_s}$ is a free scalar muitiplet in de Sitter space with mass $m^2=(2-s)(s+d-2)$, here the ``dot'' means derivative with respect to conformal time $\eta$. 
\end{appendices}

\end{document}